\documentclass[fleqn,usenatbib]{mnras}

\usepackage[T1]{fontenc}

\DeclareRobustCommand{\VAN}[3]{#2}
\let\VANthebibliography\thebibliography
\def\thebibliography{\DeclareRobustCommand{\VAN}[3]{##3}\VANthebibliography}

\usepackage{graphicx}	% Including figure files
\usepackage{epstopdf}
\DeclareGraphicsExtensions{.png,.pdf,.ps,.eps}
\usepackage{amsmath}	% Advanced maths commands
\usepackage{amssymb}	% Extra maths symbols
\usepackage{threeparttable} % Extra table options
\usepackage{multirow} % Extra table options
\usepackage{natbib} % For citations without parenthesis

\usepackage{newtxtext,newtxmath}

\title[On the single eclipse in WR22]{On the nature of the single eclipse per 80d orbit of the H-rich luminous WN star WR22 \thanks{Based on data collected by the {\em BRITE-Constellation} satellite mission, designed, built, launched, operated and supported by the Austrian Research Promotion Agency (FFG), the University of Vienna, the Technical University of Graz, the University of Innsbruck, the Canadian Space Agency (CSA), the University of Toronto Institute for Aerospace Studies (UTIAS), the Foundation for Polish Science \& Technology (FNiTP MNiSW), and National Science Centre (NCN). 
}}

\author[G. Lenoir-Craig et al.]{
G. Lenoir-Craig,$^{1}$\thanks{E-mail: guillaume.lenoir-craig@umontreal.ca}
I.I. Antokhin,$^{2}$
E.A. Antokhina,$^{2}$
N. St-Louis,$^{1}$
and A.F.J. Moffat$^{1}$
\\
$^{1}$D\'ept. de physique, Univ. de Montr\'eal, C.P. 6128, Succ. C-V, Montr\'eal, H3C 3J7, Canada, and Centre de Recherche en Astrophysique du Qu\'ebec\\
$^{2}$Moscow Lomonosov State University, Sternberg State Astronomical Institute, 119992 Universitetsky prospect, 13, Moscow, Russia\\
}

\date{Accepted XXX. Received YYY; in original form ZZZ}

\pubyear{2020}

\begin{document}
\label{firstpage}
\pagerange{\pageref{firstpage}--\pageref{lastpage}}
\maketitle

\begin{abstract}

WR22 = HD 92740 is a bright (V = 6.4 $mag$), intrinsically luminous, double-line WN7h + O9III-V binary exhibiting one sharp 8\% deep eclipse near periastron in its elliptical (e = 0.6) 80-day orbit, when the WR-star passes in front of the O star, with no secondary eclipse. We apply two models (L96, A13) to probe the optical space-based light curves from {\em BRITE-Constellation}, including three separate, complete eclipses, that show increased (o-c) scatter compared to the rest of the observations outside the eclipses, likely due to O-star light encountering WR wind-clumps. L96 is a simple atmospheric-eclipse model, often applied to close WR+O binaries, where the O-star is considered a point-source. A13 considers a finite-disk O-star and allows for atmospheric, photospheric and reflection components to the eclipse, permitting a better characterization of its shape through a more physically realistic description of the structures for both stars in WR22. Nevertheless, A13 is still susceptible to uncertainties in the luminosity of the O-star before unique values for the orbital inclination and WR mass-loss rate can be estimated. We present solutions for the two extremes of the O-star, O9V and O9III. As photometry alone cannot allow us to discriminate between these, we compared our results to the spectral models found in the literature and determined the correct solution to be O9V. Our best-fit A13 Model 1 gives $i = 83.5 \pm 0.4^{\circ}$, with $\dot M_{\rm WR} = (1.86 \pm 0.2) \times 10^{-5} \dot M_{\odot}/yr$. The flux ratio in the red {\em BRITE} band in this model is $F_{\rm O}/F_{\rm WR} = 0.064\pm 0.002$.

\end{abstract}


%%%%%%%%%%%%%%%%%%%%%%%%%%%%%%%%%%%%%%%%%%%%%%%%%%

%%%%%%%%%%%%%%%%% BODY OF PAPER %%%%%%%%%%%%%%%%%%

\section{Introduction}
When normal stars reveal eclipses, there are usually two eclipses per orbit. However, when the orbit is highly elliptical there could be only one eclipse, depending on the stellar sizes in relation to the projected (variable) separation. Clearly though, a relatively high orbital inclination is required in any case, especially for longer orbital periods.

However, in the case of stars with extended envelopes, such as WR stars with very strong winds, there is also the possibility of only an atmospheric eclipse when the WR star passes in front of its (usually) much weaker-wind O-type companion. This produces only one eclipse (via scattering by WR-wind free electrons of O-star light out of the line-of-sight) per orbit, whether circular or not (the equivalent atmospheric eclipse of WR-star light scattered off the O-star wind half an orbit later will be at least an order-of-magnitude weaker and thus of little concern). An atmospheric eclipse can also occur even for relatively low inclinations, although it may require high S/N to detect and characterize, especially for the lowest inclinations, when the eclipse amplitude is also low. 

This led \cite{Lamontagne96} to examine a fairly complete sample of short-period (P < 30 $d$) WR+O binaries known at the time, for such atmospheric eclipses. This significantly enlarged the number of binary systems with known inclination and thus allowed more masses to be obtained from radial-velocity (RV) orbits, which only yield M $\sin ^3(i)$. As an important by-product, it also allowed one to get an estimate of the WR mass-loss rate independent of clumping in the wind.

However, it was noted by \cite{Lamontagne96} that the WN7h + O9III-V binary system WR22 = HD 92740, whose long 80 $d$-period was first determined by \cite{MoffatSeggwiss78} and confirmed by \citet*{Conti79}, shows only one relatively shallow but sharp eclipse near its periastron passage in its $e = 0.6$ elliptical orbit (\cite{Balona89}; \cite{Gosset91}). There is no trace of a second eclipse half an orbit later near apastron, suggesting that the observed eclipse may be either photospheric and partial, or, since the eclipse occurred when the WR-star passed in front, leaving open the possibility of a purely atmospheric instead of a more general combined photospheric/atmospheric eclipse.

WR22 is an important system, containing a highly luminous, H-rich WN star in a long-period binary. The two other known bright WR stars in the Carina Nebula are also luminous H-rich WN stars, with WR24 likely being single and WR25 being a 208d binary \citep{Gamen2006}, though almost two visual magnitudes fainter than WR22 and WR24 due to higher interstellar extinction. This makes WR22 a very useful target, the apparently brightest binary among known WNLh stars, along with the bonus of being a eclipsing system.

In this study, we explore the nature of WR22's eclipse in more detail, thanks to the ability of the nanosatellites of the {\em BRITE-Constellation} to intensely monitor bright stars in precision optical photometry for up to half a year non-stop. Previous ground-based observations of WR22 show only marginal coverage of this eclipse due to both the long orbital period and the short (but not short enough for proper coverage from one ground-based site) three-day duration of the eclipse.

\section{Observations}

\begin{figure*}
\makebox[\textwidth][c]{%
  \includegraphics[width=1.2\textwidth]{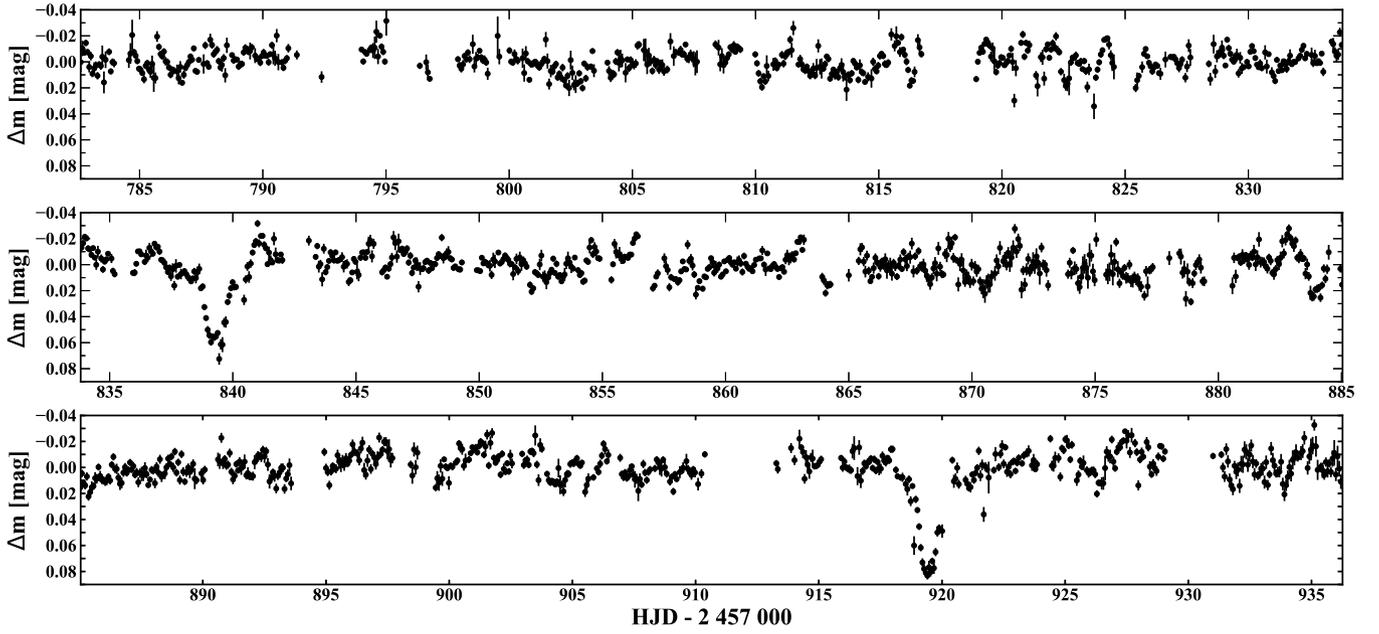}%
}
\caption{De-correlated time-dependent light curve of WR22 obtained by the {\em BRITE-Heweliusz} satellite in it's 2017 coverage of the Carina field, after subtracting off the mean. The two-sigma error bars are based (mainly) on the instrumental scatter during a BRITE orbit (see text). The two eclipses separated by $\sim$80 $d$ are obvious.}
\label{fig:Figure 1}
\end{figure*}

\begin{figure*}
\makebox[\textwidth][c]{%
  \includegraphics[width=1.2\textwidth]{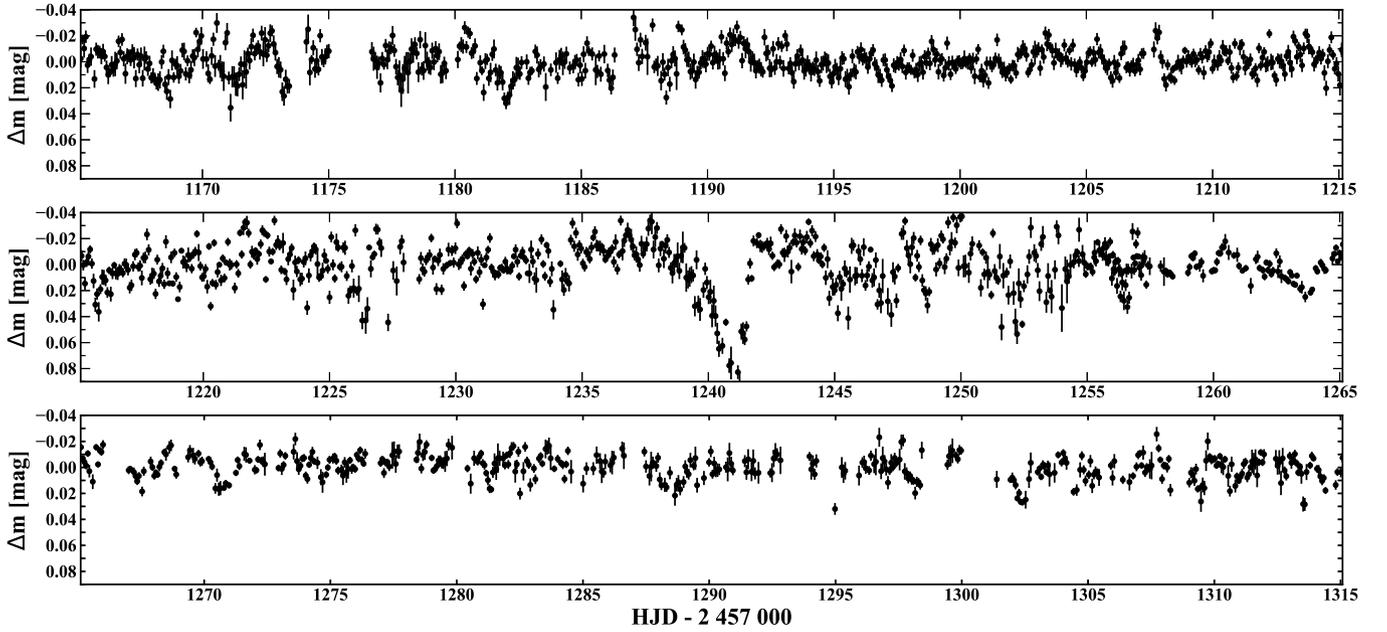}%
}
\caption{As in Fig. \ref{fig:Figure 1} except for 2018. The single eclipse observed near the middle of the run is obvious.}
\label{fig:Figure 2}
\end{figure*}

%The next two paragraphs were lifted from Ramiaramanantsoa et al. 2018 on zeta Pup and revised accordingly.
{\em BRITE-Constellation} (\citealt{Weiss14}; \citealt{Pablo16}) consists of a network of five nano-satellites each housing a 35-$mm$ format KAI-11002M CCD imaging detector fed by a 30-$mm$ diameter f/2.3 telescope through either a blue (b) filter ($390 - 460$ $nm$) or a red (r) filter ($545 - 695$ $nm$): {\em BRITE-Austria} ({\em BAb}), {\em Uni-BRITE} ({\em UBr}), {\em BRITE-Heweliusz} ({\em BHr}), {\em BRITE-Lem} ({\em BLb}) and {\em BRITE-Toronto} ({\em BTr}). All the satellites were launched into low-earth orbits of period $\sim 100$ $min$, and are now fully operational. With a $\sim 24^\circ \times 20^\circ$ effective field of view, each component of {\em BRITE-Constellation} performs simultaneous monitoring of 15 -- 30 stars brighter than $V \sim 6$ $mag$. A given field is observed typically over a $\sim$ 6-month time base.

WR22 was monitored by {\em BHr} in the {\em BRITE} Carinae field no. 24 (2017 Jan 10 - July 12) and no. 36 (2018 Feb 15 - July 15). The {\em BRITE} detector pixel-size is  27.3", with a resolution of FWHM $\sim$5 pixels and thus comfortably including, yet isolating, the light from WR22. Short 4-5 $s$ exposures were taken at a median cadence of 20 $s$ during $\sim$1 -- 30\% of each
$\sim$ $100-min$ {\em BRITE} orbit, the remaining time unused due to stray light interference, blocking by the Earth, and limited data-download capacity. 

All observations were performed
in chopping mode (\cite{Pablo16}; \cite{Popowicz16}; \cite{Popowicz17}. Raw light curves were extracted using the reduction
pipeline for {\em BRITE} data which also includes corrections
for intra-pixel sensitivity \citep{Popowicz16}. Then post-extraction
decorrelations with respect to instrumental effects
due to CCD temperature variations, centroid position and
satellite orbital phase were performed on each observational setup for
each satellite according to the method described by \cite{Pigulski16}. 

In the resulting final decorrelated light curves, we see no obvious and significant variations on timescales less than the short sampling time during each {\em BRITE} orbital period, that could
be qualified as intrinsic to the star rather than pure instrumental
noise. Therefore in order to gain in precision, we calculated
satellite-orbital mean fluxes and their uncertainties to create the final light curves in the optical red band. This would also essentially eliminate any linear trends, if any should prevail during a BRITE orbit. We also performed
removal of outliers during the decorrelation process,
such that it is reasonable to adopt orbital mean fluxes instead of median fluxes.

%[?] Then, to extract the orbital means, we tested two different methods : (1) a simple average of the flux values taken within an orbit, and (2) an average taken to be the mid-point of a linear fit of the flux values within an orbit. [OK? -->] We noticed no significant difference between the resulting root mean square (rms) values of the mean standard deviations obtained from the two methods, the second one being only slightly better. Therefore we adopted the second method to generate the final binned light curves in the two filter bands which we use to extract information on the intrinsic variability of the star (Figure 1).

\section{light curve analysis}

\subsection{General}

\begin{figure}
\makebox[\columnwidth][c]{%
  \includegraphics[width=1.2\columnwidth]{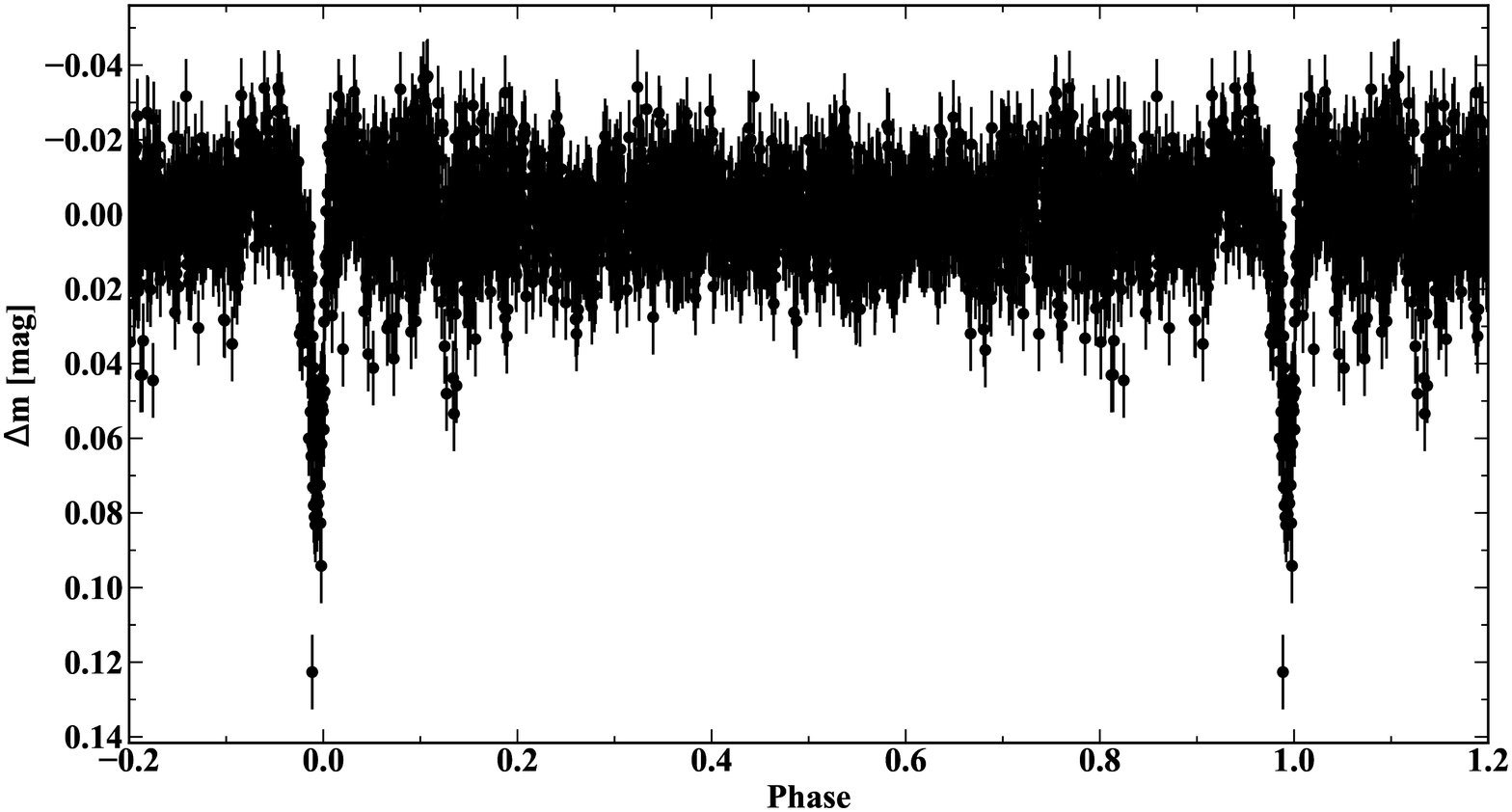}%
}
\caption{Complete phased light curve of WR22 for 2017 and 2018 combined from the BHr satellite.}
\label{fig:Figure 3}
\end{figure}

We show the reduced {\em BRITE} light curves from the 2017 and 2018 runs in Figs. \ref{fig:Figure 1} and \ref{fig:Figure 2}, respectively. The de-trended data presented in these figures are available online: see the Data Availability section at the end of this article. A total of three eclipses were covered, basically identical except for the intrinsic noise pattern mainly of the much brighter WN component, which will be examined in a separate study (Lenoir-Craig et al., submitted) after subtracting off the best-fitted eclipse. The data from the 2018 run appear to show some additional instrumental scatter associated with the general and gradual degradation of the detector. Since we have no way of separating this from the intrinsic noise mainly of the WN component (which is generally much larger than the instrumental noise), we ignore this difference. Fig. \ref{fig:Figure 3} shows a combined phased light curve using the ephemeris in Table \ref{Table 1}.

\begin{figure}
\includegraphics[width=\columnwidth,trim=1cm 0.0cm 3.0cm 2.0cm,clip=True]{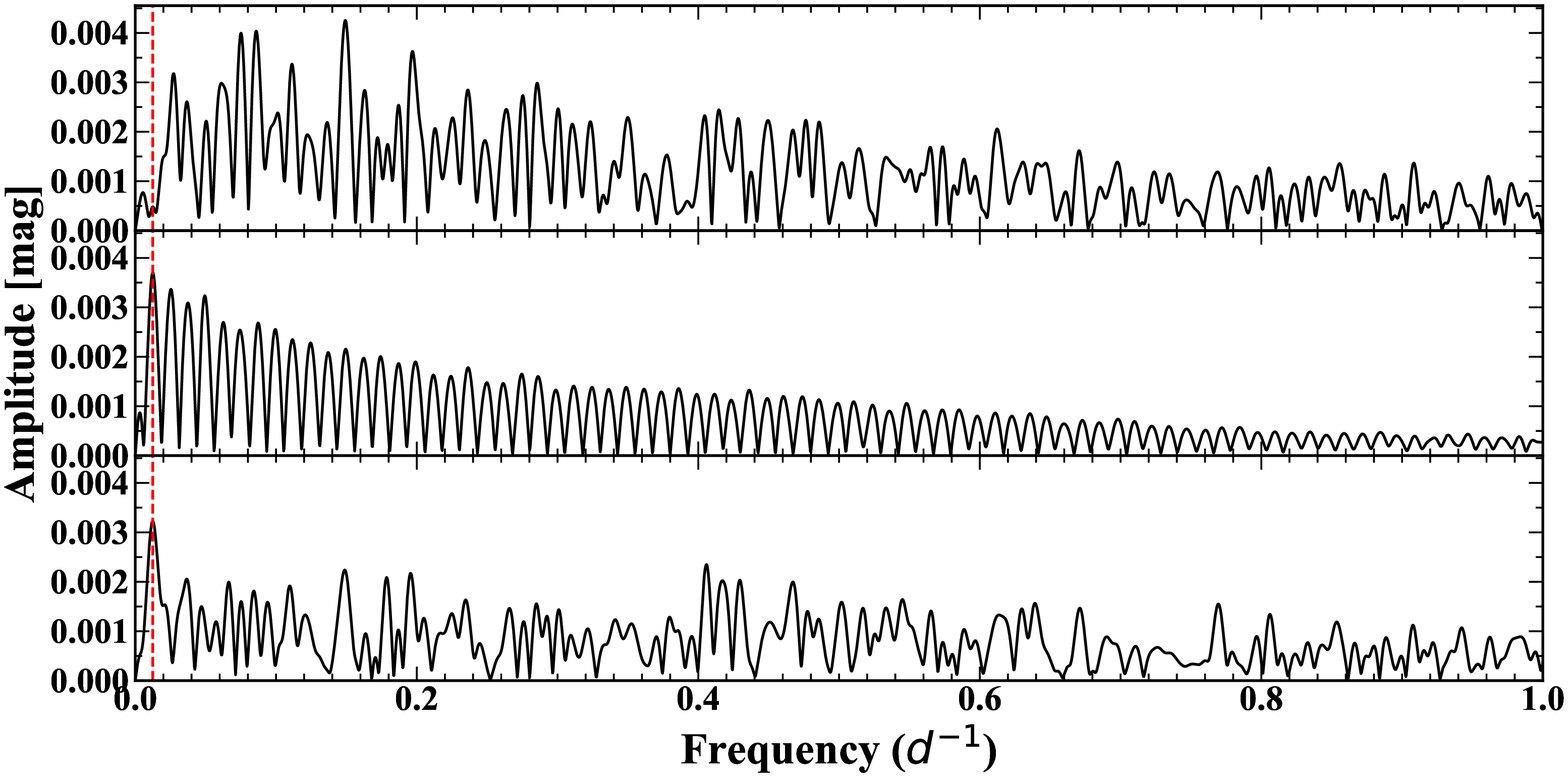}
\caption{Top: periodogram of the light curve presented in Fig. \ref{fig:Figure 1} for 2017. Middle: periodogram of the fitted L96 eclipse model without the data but following the same sampling as the data. The high number of oscillating harmonics is due to the presence of two eclipses in that time-series, which interfere with one another in the Fourier domain. Bottom: periodogram of the 2017 observations after subtracting the fitted L96 model (see text). The frequency position of the $80.336$ $d$ orbital period is marked with a red vertical dashed-line in all panels.}
\label{fig:Figure 4}
\end{figure}

\begin{figure}
\includegraphics[width=\columnwidth,trim=1cm 0.0cm 3.0cm 2.0cm,clip=True]{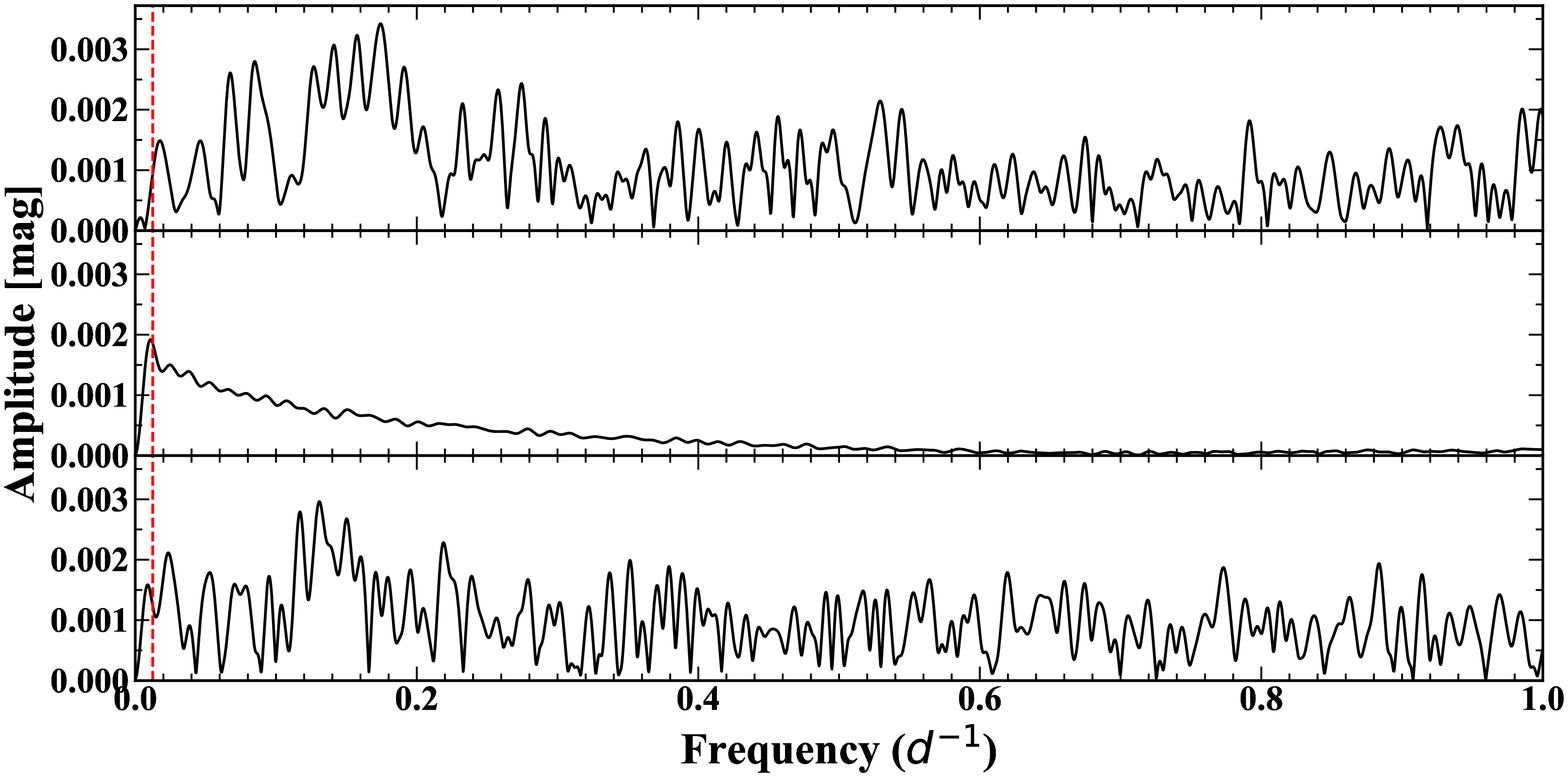}
\caption{Top: periodogram of the light curve presented in Fig. \ref{fig:Figure 2} for 2018. Middle: periodogram of the fitted L96 eclipse model without the data but following the same sampling as the data. This time, the harmonic spectrum is still seen, albeit at somewhat lower amplitude but without the extreme oscillations, due to the presence of only one narrow eclipse. Bottom: periodogram of the 2018 observations after subtracting the fitted eclipse model. As in Fig. \ref{fig:Figure 4}, the $80.336$ $d$ orbital period is marked with a red vertical dashed-line in all panels.}
\label{fig:Figure 5}
\end{figure}

Figs. \ref{fig:Figure 4} and \ref{fig:Figure 5} show periodograms of the 2017 and 2018 data, respectively. These include for the observed data (upper panel), for the best atmospheric model eclipse (see below) without data but for the same distribution of data-points (middle) and for after subtraction of the model eclipse (bottom). In the periodogram of the observed data from 2017 (upper panel of Fig. \ref{fig:Figure 4}), the peak corresponding to the orbital period is effectively suppressed by noise interference from the rest of the light curve (and no such significant peak was expected in the 2018 observed data periodogram since only one eclipse was observed then). The observed periodograms obtained after subtracting off the eclipse are dominated by a forest of  low-frequency peaks reflecting the stochastic nature of the variability that likely arises from random clumping in the WR wind (as, for example, in the WN8h star WR40: \cite{Ramia19}), whose stochastic nature manifests itself by the lack of a match in the detailed power peaks for each run. The 2018 periodogram also appears somewhat different in overall nature with lower density of low-frequency power peaks, probably a result of what seems to be increased instrumental noise in the 2018 light curve in Fig. \ref{fig:Figure 2}. The difference in the model eclipse periodogram for each run reflects the fact that there are two eclipses in 2017 and only one eclipse in 2018, leading to a higher amplitude and clearer harmonics in the former, due mainly to the interfering effects of the two narrow eclipses. Note that the models in this paper assume smooth winds, with no allowance for stochastic wind-clumping.

We then used the well-known ephemeris of WR22's binary parameters (\cite{Rauw96}; \cite{Schwei99}, with priority for the latter, which are generally more precise) to combine these three eclipses into one phased light curve, as already shown in Fig. \ref{fig:Figure 3}. Besides the eclipse itself, the stochastic nature of the rest of the light curve is obvious, with a total spread of $\sim 0.06$ $mag$.

%From Figure \ref{fig:Figure 4} [<-- original Fig.4 is removed, since such different curves for different i are already illustrated in L96 Figs. 2 and 3. Note that Mdot is only a multiplicative factor affecting only the eclipse depth once i is chosen. - TM.] 
The basic question arises whether the eclipse is photospheric or (purely) atmospheric. In the former case (which can also include an atmospheric component), if the orbit were circular and the orbital inclination were high enough, one is likely to see two eclipses, one at either conjunction as each star in turn eclipses the other. But in an elliptical orbit as here for WR22, it is possible that one could only be seeing one eclipse near periastron when the two stars are closer together, while the second eclipse is hindered by a larger orbital separation towards apastron. In the case of an atmospheric eclipse, the question of seeing more than one detectable eclipse is irrelevant when the companion to the WR star is a weak-wind O-star, as is the case here. As for the nature of the eclipse, we first explore the possibility of an atmospheric eclipse, mainly because it is so simple and easy to apply. Another motivation for using this model was that it had been applied to a number of WR+O binaries with small inclination angles. Thus it would be interesting to see how well (or badly) this model performs while being applied deliberately to a binary system with with a large orbital inclination. After this we will explore the much more complex, but physically more rigorous, model allowing for both photospheric and atmospheric extinction, which is physically more appropriate. In both models, the wind of the WR component is assumed to be spherically symmetric. Although there is a wind-wind collision in the system, the contact surface is located very close to the O star or even collapses onto its surface at periastron (see \cite{Parkin11} for detailed hydrodynamical simulations). Thus, at orbital phases close to periastron (where the eclipse occurs), the line of sight from the O star to the observer (along which the decrease of its flux occurs) passes through the unperturbed spherically symmetric part of the WR wind.

\subsection{Analysis as an atmospheric eclipse, treating the O-star as a point-source}

We follow the simple recipe for fitting an atmospheric eclipse developed by \cite{Lamontagne96} for short-period WR + O binary systems with circular orbits (L96 model), with minor modifications, the most important of which being the removal of the optically-thin wind approximation. We adapt this to the longer elliptical orbit of WR22 using the orbital elements of \cite{Schwei99}, slightly preferred over the earlier work of \cite{Rauw96}. 

In the L96 model with or without the optically thin wind approximation, the O star is assumed for simplicity to be a point-source, with flux variations due to extinction caused by single scattering off free electrons in the WR wind. This model does not include the possibility of a geometrical eclipse of the O star by the WR disk, which we add here, even if it may be somewhat artificial due to the assumption of a point-source O-star. The optical depth is an integral along the line of sight from the O star to the observer (and becomes less accurate for high inclinations where these simplifications no longer apply):

\begin{equation}
  \tau(\phi) = \tau_0\int\limits_{z_0/a}^\infty \frac{d(z/a)}{(r/a)^2V(r)/V_\infty} ,
  \label{eq:1}
\end{equation}

\noindent where

\begin{equation}
  \begin{array}{rcl}
   (r/a)^2 & = & ( d/a \cos(i)\cos(v+\omega-\pi/2) )^2\\
   & & + ( d/a \sin(v+\omega-\pi/2) )^2\\
   &  & +(z/a)^2 ,
  \end{array}
  \label{eq:2}
\end{equation}

\noindent $d$ is the current distance between the components, $\omega$ is the longitude of periastron of the O star, and $v$ is the true anomaly,

\begin{equation}
  z_0/a = -d/a \sin(i) \cos(v+\omega-\pi/2) .
  \label{eq:3}
\end{equation}

\noindent The wind velocity $V(r)$ follows the usual $\beta$-law

\begin{equation}
  V(r) = V_\infty\left(1-\frac{R_{\rm WR}}{r}\right)^\beta ,
  \label{eq:4}
\end{equation}

\noindent Also %$\tau_0$ is 

\begin{equation}
  \tau_0 = \sigma_T n_0 a ,
  \label{eq:5}
\end{equation}

\noindent where $\sigma_T$ is the Thomson cross-section, $a$ is the semi major-axis, and $n_0$ is a fiducial electron density, with

\begin{equation}
  n_0 = \frac{\dot{M}}{4\pi m_p\mu_e a^2 V_\infty},
  \label{eq:6}
\end{equation}

\noindent where $\dot{M}$ is the WR mass-loss rate, $m_p$ is the proton mass, $\mu_e\simeq 2/(1+X)$ is the mean electron molecular weight in the WR wind and $X$ is the H fraction.

The observed light curve is fitted by

\begin{equation}
  \Delta m(\phi) = \Delta m_0 -2.5\log\frac{1+F_re^{-\tau(\phi)}}{1+F_re^{-\tau(\phi_0)}} ,
  \label{eq:7}
\end{equation}

\noindent where $\Delta m_0$ is the arbitrary magnitude zero-point; %accounting for the magnitude difference between the target and the comparison star
$F_r=F_{\rm O}/F_{\rm WR}$ is the O/WR flux ratio; $\phi_0$ is the reference phase such that the relative magnitude (the second term in eq. \ref{eq:7}) at this phase is equal to 0; and the solution for $\tau$ was taken from $B = (\Delta m - \Delta m_0)/A$ in eq. 13 of \cite{Lamontagne96} for $\beta = 1$, leading to $\tau(\phi) = a \tau_0 B$.

While computing a model light curve, at each orbital phase we verify the condition for a total eclipse (the sky plane projected distance between the O star, assumed to be a point-source, and the center of the WR star is smaller than the WR radius). Thus our model can in principle be applied to the case of a total (geometrical) eclipse. However, as such an eclipse is clearly not observed in WR22 (none of our three observed eclipses has a flat bottom), we restrict the range of possible inclination angles so that the minimal projected distance is equal to $R_{\rm WR}$. This maximal allowable inclination is defined by the simple formula

\begin{equation}
  \tan(i) \leq \frac{a\sin(i)(1-e^2)}{(1+e\cos(v_{\rm conj}))R_{\rm WR}} ,
  \label{eq:8}
\end{equation}

\noindent where $v_{\rm conj}=\pi/2-\omega$ is the true anomaly at superior conjunction of the O star.

The L96 model has five free parameters:  $F_r$, $i$, $\tau_0$, $\Delta m_0$ and $\Delta\phi$. We consider $F_r$ as a free parameter, since the fixed flux ratio of \cite{Schwei99} was obtained assuming a total eclipse, which is not a valid assumption here.

Fitting $\Delta\phi$ is necessary since the ephemeris from \cite{Schwei99} corresponds to periastron passage, and with their eccentricity and argument of periapsis values, the eclipse should occur at phase +0.001, which is not the case as can be seen in Fig. \ref{Figure 7} where the eclipse bottom is at phase -0.0075. This discrepancy is possibly related to an inaccuracy in their $T_0$ measurement, their orbital period calculation, or both. $\Delta\phi$ is searched for by translating the phased light curve (with phases calculated using the \cite{Schwei99} ephemeris) during the fitting process, and the resulting phase shift is then added to all the data-points in the light curve, allowing us to correct for the discrepancy.

Some of the free parameters may correlate with each other, most notably $F_r$, $i$ and $\tau_0$. To avoid the risk of trapping the minimization algorithm in a spurious local minimum, we chose to run the algorithm on a grid of $F_r$ to fully explore $\chi^2$ as its goodness-of-fit function.

The photometric errors (see the error bars in Figs. 1 and 2) are much smaller than the actual scatter even out of eclipse and the empirical distribution of data points is not Gaussian. This is due to the random nature of WR-wind clumps that create light changes that are coherent on time-scales of hours rather than all points being completely independent from each other. However, the overall number of data points in the part of the light curve which is far from the deviating influence of the eclipse and its wings is very large compared to the other parts, and hence includes a large enough number of coherent substructures that are stochastic and independent from each other, so that deviations of the empirical distribution  from a Gaussian distribution are relatively small. For this reason we adopted $\sigma = 0.0105$ calculated between phases 0.1 and 0.9 as the basic empirical data scatter for the calculation of the reduced $\chi^2$, preferred over the scatter calculated over the whole light curve of $\sigma = 0.0131$. This interval was chosen as it is far from the eclipse and its wings, where the sigma values would be larger. 

Using the proper beta-law to describe the velocity profile of the WR wind is also important, since the eclipse is very narrow in orbital phase and thus includes part of the WR wind near its base, where the choice of the beta exponent becomes more critical. As in other H-rich WNL stars and some extreme Of stars, we take beta  $\sim$1 \citep{LepineMoffat08}, as opposed to larger beta-values in classical WR winds with higher density \citep{LepineMoffat99}.

The fitting procedure begins with the definition of the fixed stellar radius ($R_{\star} = 22.65 R_{\odot}$) and $X_H = 0.44$ for the WR star, both from \cite{Hamann19}. We then define a grid of flux ratios ($F_r$) to be tested. For each flux ratio value, the four remaining free model parameters are adjusted using a non-linear least-squares algorithm minimizing the $\chi^{2}$ to reach the required light curve shape.

\begin{figure*}
\includegraphics[width=\textwidth]{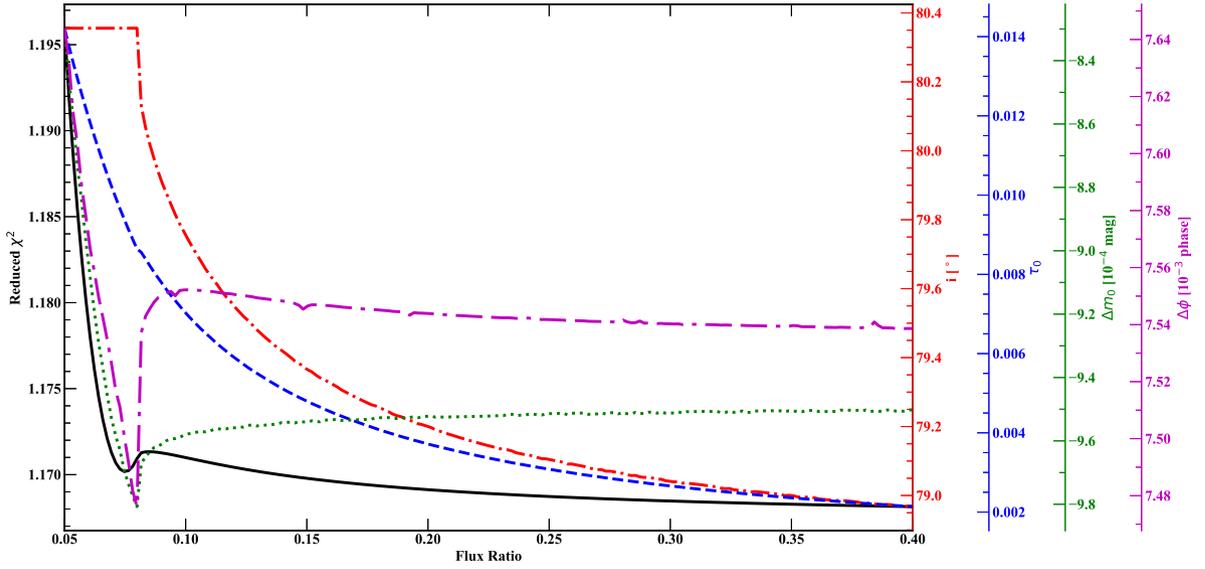}
\caption{The four free parameters of the L96 model, along with reduced $\chi^2$ from the fit, varying with the flux ratio $F_r = F_{\rm O}/F_{\rm WR}$. The solid black, dash-dotted red, dashed blue, dotted green and dash-dash-dotted magenta curves respectively correspond to the reduced $\chi^2$, the inclination angle $i$ and the parameters $\tau_0$, $\Delta m_0$ and $\Delta\phi$.}
\label{Figure 6}
\end{figure*}

\begin{figure}
\includegraphics[width=\columnwidth,trim=0.5cm 0.0cm 3.0cm 2.0cm,clip=True]{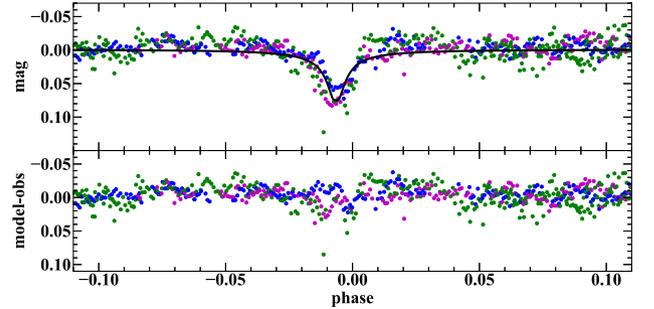}
\caption{Top: Phased light curve of WR22's eclipses from 2017 and 2018, fitted in black (for the whole light curve, not just the part shown here) with the \protect\cite{Lamontagne96} equation for a WR wind with $\beta=1$. Blue dots correspond to the data acquired around the time of the first observed eclipse in 2017, magenta dots are for the 2017 second eclipse and green dots for the 2018 only eclipse. Bottom: Residuals obtained from subtracting the fit from the data. These show no obvious deviating trends.}
\label{Figure 7}
\end{figure}

Fig. \ref{Figure 6} shows how the free L96 parameters vary with the flux ratio, along with the reduced $\chi^{2}$ from the fit. At small flux ratio, in order to respect the required eclipse depth, the model needs to increase $i$ and decrease $\tau_0$, while $\Delta m_0$ is adjusted to keep the out-of-eclipse part of the model aligned with the data and $\Delta\phi$ varies to keep the lowest part of the model centered on the eclipse minimum. However, since $i$ has a maximum value of 80.36$^\circ$ (above that value, the point-source O star goes behind the WR disk and the eclipse becomes total), only $\tau_0$ can be adjusted until an $F_r$ value of $\sim 0.08$ is reached. Past this point, the projected point-source O star can move away from the WR disk and both $i$ and $\tau_0$ decrease to compensate for the rising $F_r$. The local minimum in the reduced $\chi^{2}$ at $F_r = 0.073$ is within the error bars of the spectrum-based value of the flux ratio of $F_r = 0.13 \pm 0.12$ calculated when adopting the typical visual absolute magnitudes for WN7h and O9V stars shown in Table \ref{Table 1}, but well outside the flux ratio $F_r = 0.24 \pm 0.06$ calculated when using the typical visual absolute magnitude for O9III stars. The asymptotic behavior of the $\chi^{2}$ at large $F_r$ can be explained by taking the second order Taylor expansion of eq. \ref{eq:7} applied to the phase of eclipse minimum, resulting in the expression below for the eclipse depth:

\begin{equation}
  depth \approx \frac{\tau_{\phi_0}}{(1+1/F_r)},
  \label{eq:9}
\end{equation}

\noindent where at large $F_r$, the $1/F_r$ term becomes very small and $depth \approx \tau_{\phi_0} \approx 0.08$, thus becoming independent of $F_r$. Hence this is why, even if the reduced $\chi^{2}$ of the model follows a decreasing trend with increasing flux ratio past $F_r \sim 0.08$ and reaches lower values than that of the local minimum, we choose to adopt the parameter values corresponding to the local minimum in $\chi^{2}$ at $F_r = 0.073$, as the corresponding value of flux ratio is more coherent with previous estimations of the contribution of the O-component to the total flux of the system (\citealt{Rauw96} \& \citealt{Hamann19}).

\begin{table}
\caption{Parameters of the adopted L96 light curve solution.}
\label{Table 1}
\begin{threeparttable}
\centering
\begin{tabular}{c c} 
Parameter & Value \\
\hline
\multicolumn{2}{c}{Assumed Parameters} \\
\hline
P [days] & {80.336 $\pm$ 0.0013 \tnote{a}}\\
$T_0$ [HJD] & {2 450 126.97 $\pm$ 0.14 \tnote{a}}\\
$R_{\rm WR}$ [R$_{\odot}$] & 22.65 \tnote{b} \\
$X_H$ & 0.44 \tnote{b} \\
e & {0.598 $\pm$ 0.010 \tnote{a}}\\
$\omega_{\rm O}$ [$^{\circ}$] & {88.2 $\pm$ 1.6 \tnote{a}}\\
a$_{\rm WR}\sin{i}$ [$10^{6}$ km] & {62.5 $\pm$ 0.9  \tnote{a}} \\
a$_{\rm O}\sin{i}$ [$10^{6}$ km] & 168.2 $\pm$ 9.0  \tnote{a} \\
M$_{v}^{\rm WR}$ & -6.8 \tnote{b} \\
M$_{v}^{\rm O9V}$ & -4.0 \tnote{c} \\
M$_{v}^{\rm O9III}$ & -5.3 \tnote{c} \\
v$_{\infty}^{\rm WR}$ [km/s] & {1785 \tnote{b}} \\
$\beta$ & 1 \tnote{d} \\
$\mu_{e}$ & 1.39 \tnote{b} \\
%$\chi^2$/d.o.f & 4451.88/3442 = 1.293 \\
\hline
\multicolumn{2}{c}{Fitted Parameters} \\
\hline
$F_{\rm O}/F_{\rm WR}$ & $0.073$ (the grid point of the best-fit model)\\
$\tau_0$ & 0.0093 $\pm$ 0.0006 \\
$\Delta m_0$ [mmag] & -0.00097 $\pm$ 0.00002 \\
i [$^\circ$] & 80.36 $\pm$ 0.08 \\
$\Delta\phi$ [phase] & -0.0085 $\pm$ 0.0003 \\\
$\chi^2$/d.o.f & 4027/3442 = 1.17 \\
\hline
\multicolumn{2}{c}{Derived Parameter} \\
\hline
$\dot M_{\rm WR}$ [$10^{-5}\rm M_\odot$\ / year] & 2.6 $\pm$ 0.2 \\ 
\hline
\end{tabular}

\begin{tablenotes}
\item[a] \cite{Schwei99}; O9V preferred over O9III
\item[b] \cite{Hamann19}
\item[c] \cite{Martins05}
\item[d] Wind velocity-law exponent
%\item[e] \cite{Martins05}
\end{tablenotes}

\end{threeparttable}
\end{table}

%Recall that the actual scatter of the observed data points outside eclipse is much larger than the photometric uncertainties of individual points. The pattern of the  variability outside of eclipse seems to be not completely random and may represent some kind of auto-regressive process, e.g. short-term coherency due to clumps coming and going on timescales of $\sim$10 hours. This variability will also be the subject of a separate paper. The residual scatter during the eclipse ($\sigma_ecl = 0.0176$) is much greater the residual scatter out of the eclipse ($\sigma_outecl = 0.0103$) (as also found in applying the photospheric model below), allowing us to conclude that the atmospheric fit is mathematically acceptable and making it inappropriate to assign the out-of-eclipse uncertainty to the data points during the eclipse. 

Of particular interest in the parameter summary of Table 1 are the orbital inclination $i = 80.36 \pm 0.08^\circ$ and the WR mass-loss rate $\dot M_{\rm WR} = (2.6 \pm 0.2)10^{-5} M_\odot\ / yr$. While the latter is normal for luminous WN stars with hydrogen, being within the average $\dot M$ range for WN7h stars of $\dot M_{\rm WR} = (2.8 \pm 0.8)10^{-5} M_\odot\ / yr$ from \cite{Hamann19}, the former could have a problem with the assumptions in applying the \cite{Lamontagne96} model, even if the fit is formally acceptable. In that model, it is assumed that the O-star is a point-source, which is ultimately untenable. %This approximation leads to a model eclipse too narrow to appropriately match the data, leading to the $\chi^2 / d.o.f. = 1.2934$ value for the best fit, which informs us that the model is statistically rejected since the corresponding significance level of $ 1.25 \times 10^{-28}$ is very close to zero. %However, the $\chi^2$ value was computed assuming a normal distribution of data-point errors. Recall that the actual scatter of the observed data points outside eclipse is much larger than the photometric uncertainties of individual points. The pattern of the  variability outside of eclipse seems to be not completely random and may represent some kind of auto-regressive process. This variability will be the subject of a separate paper. For the purpose of our modelling we considered it as random Gaussian noise and estimated the empirical data-point error by computing the uncertainty $\sigma_{obs}=$  in the phase interval $0.1-0.9$. We then assigned this uncertainty to all data points.

%Nevertheless, the fit looks good, although the question remains whether it provides realistic parameters. 
In the next section, we explore models allowing for a finite disk for the O-star plus scattering in the WR wind, thus enabling the eclipse to have a photospheric component, which is a much more complicated and difficult case but a better approximation. Once this is done, we compare the light curve solutions in Section \ref{section:Discussion}. 

\subsection{Analysis using a Roche model plus wind}

We can gain an idea for the need of applying photospheric as opposed to only atmospheric eclipses by simply examining the conditions for photospheric eclipses, i.e. no eclipse if \mbox{$r\cos(i) > R_{\rm WR} + R_{\rm O}$}, where $r$ is the separation between the centres of the two stars at the respective conjunctions and $R_{\rm WR}$, $R_{\rm O}$ are the photospheric radii of the stars.

We list in Table \ref{Table:2} the pertinent parameters during both conjunctions (almost exactly as it turns out) at periastron (where we know that one eclipse does occur) and apastron (where a second eclipse might have occurred), based on the orbits of \cite{Rauw96} and \cite{Schwei99}. For the WR star we take $R_{\star}$ and $R_{2/3}$ as extreme values between the hydrostatic stellar radius \citep{Hamann19} and the pseudo-photosphere in the wind \citep{Schwei99}. For the O-star we take mean radii for O9V stars and O9III stars \citep{Martins05}.

\begin{table}
\caption{Parameters that determine if there are photospheric eclipses. All distances and radii are in $R_\odot$. The conjunctions are those of the O star. $i$ is assumed to be $80.36^\circ$, the orbital inclination value of the adopted L96 light curve solution.}
\label{Table:2}
\begin{threeparttable}
\centering
\begin{tabular}{c c c} 
Parameter & \cite{Rauw96} & \cite{Schwei99} \\
\hline
e             &          0.559 $\pm$ 0.009    &   0.598 $\pm$ 0.010\\
a\,$\sin{i}$     &    361.1 $\pm$ 14.4     &   331.5 $\pm$ 12.9\\
d$_{\rm conj}^{\rm sup}\cos{i}$       &      24.83        &       22.64\\
d$_{\rm conj}^{\rm inf}\cos{i}$      &      87.72        &       89.91\\
\hline
 & Minimum & Maximum \\
\hline
R$_{\rm WR}$ & 22.65($R_\star$) & 28.5($R_{2/3}$) \\
R$_{\rm O}$  & 7.53(V) & 13.38(III) \\
R$_{\rm WR}$ + R$_{\rm O}$ & 30.38 & 42.2 \\
\hline
\end{tabular}
\end{threeparttable}
\end{table}

From Table \ref{Table:2}, we see that, indeed, a photospheric eclipse is likely at/near periastron and unlikely at/near apastron. In order to explore the former case, the WR22 light curve is fitted in this section with the A13 model of \cite{ant13}, extending the standard Roche lobe model by including a wind around the WR component and assuming the wind of the O companion to be negligibly weak by comparison. The model is based on a computer code which allows one to calculate light and radial velocity curves, as described by \cite{ant88, ant96} and \cite{ant00}. The code is similar to that of \cite{wils71} and \cite{wils79} and has been applied to binary systems of various types, enabling the computation of light and radial velocity curves for circular or eccentric orbits.

The shapes of both components in A13 are computed according to Roche geometry. The model subdivides the surfaces of both components into small areas and computes the flux of each area while accounting for limb and gravitational darkening, as well as mutual irradiation (reflection effects). The wind of the WR component is considered to be spherically symmetric, with a radial velocity distribution corresponding to the velocity law described in eq. \ref{eq:4}. $R_{\rm WR}$ in eq. (\ref{eq:4}) is assumed to be equal to the radius of a sphere with an equivalent volume to the one of the WR body computed with the Roche geometry. The optical depth of the wind is computed for each elementary area of the O star by numerical integration of eq. (\ref{eq:1}). The program computes monochromatic fluxes at the central wavelength of the observational pass-band.

Other methods for solving light curves of WR+O binaries have been used in the past. For example, \cite{perrier09} proposed a method based on the use of the empirical moments of the light curve, which are integral transforms evaluated from the observed time-series. The model moments are computed using very simple analytical expressions for limb darkening and wind transparency, which are then compared to the observations. The method assumes spherical symmetry for both stars and does not account for any reflection effect. To determine the observational moments, a smoothed observed light curve is required, which is achieved by spline approximation. This method can thus be applied to well-defined light curves, for which the photometric errors are much smaller than the eclipse depth, so that the spline smoothing of the observed light curve produces an unambiguous result. Unfortunately, this is not the case for the {\em BRITE} data of WR22 used in this paper (see below).

Another method for solving light curves of WR+O binaries was proposed by \cite{cher84} and further developed by \cite{ant12, ant16}. By directly solving integral equations describing the light curve via regularizing algorithms, this method has the advantage of not requiring a parametric description of the limb darkening, the wind velocity law, etc. For example, the minimum a priori constraint on the distribution of the wind transparency function (sufficient to get a stable solution) is that this function is non-negative and does not increase with increasing impact distance. However, this method also ignores the tidal distortion of the stellar shapes and the reflection effects. In addition, to obtain the transparency function, it is required that the orbital inclination angle be large enough that during the eclipse the disk of the O star overlaps the center of the WR disk. Thus, both methods are not very well suited for solving the light curve of WR22.

The input A13 model parameters are as follows:

\begin{enumerate}
  \item $P$ -- orbital period.
  \item $T_0$ -- time of periastron.
  \item $e$ -- eccentricity.
  \item $\omega$ -- longitude of periastron for the O star.
  \item $i$ - orbital inclination.
   \item $M_1\sin^3i$, $M_2\sin^3i$ -- masses of the components multiplied by $\sin^3i$. These are the observed parameters usually available from the radial velocities. $M_1$ and $M_2$ are then computed for a given $i$. 
  \item $\mu_1$, $\mu_2$ -- Roche lobe filling factors, $\mu=R/R_c$, where $R$ is the polar radius of a stellar body and $R_c$ is the polar radius of the critical Roche lobe at periastron ($0 < \mu \leq 1$). At other orbital phases, $\mu_1$, $\mu_2$ are recomputed from the condition that the volumes of stellar bodies are constant.
  \item $T_1$, $T_2$ -- stellar temperatures. Note that as the program computes the monochromatic flux assuming a black-body spectrum, these temperatures are essentially the flux scaling parameters and may not directly correspond to the actual stellar temperatures, especially for the WR star.
  \item $F_1$, $F_2$ -- ratios of surface rotation to synchronous rate.
  \item $\beta_1$, $\beta_2$ -- gravity-darkening coefficients.
  \item $A_1$, $A_2$ -- bolometric albedos.
  \item $(x,y)_1$, $(x,y)_2$ -- limb-darkening coefficients.
  \item $\lambda$ -- effective wavelength of the monochromatic light curve.
  \item $\Delta m_0$ -- the zero point of stellar magnitudes as in the L96 model.
  \item $\Delta\phi$ -- the phase shift of the observed light curve (observed minus predicted phase) due to the inaccuracy of $T_0$ and/or $P$ and/or period change.
  \item $\beta$ -- index parameter of the $\beta$-law. 
  \item $V_\infty$ -- terminal velocity of the WR wind.
  \item $\mu_e$ -- mean electron molecular weight of the WR wind.
  \item $\dot{M}$ -- WR mass-loss rate.
\end{enumerate}

Note that the last three parameters are only used when solving the direct problem, i.e. when computing the model light curve for a given set of model parameters. These parameters are not independent, being interrelated by eqs. (\ref{eq:5}) and (\ref{eq:6}). When carrying out the fitting procedure, the actual model parameter is $\tau_0$. Once it is found, the value of $\dot{M}$ can then be derived by assuming values for $V_\infty$ and $\mu_e$.

Several model parameters are either known from previous studies or can be fixed to reasonable values. $P$, $T_0$ (the periastron passage date), $e$, $\omega$, $M_1\sin^3i$, $M_2\sin^3i$ were taken from \cite{Schwei99}. The rotation of both components was assumed synchronous so the $F_{1,2}$ values were set to unity. The gravity darkening coefficients $\beta_{1,2}=0.25$ were set according to \cite{zeip24}. The non-linear ``square-root'' limb darkening for both components was computed according to \cite{hamme93}. Note that for the WR star the limb darkening is formal. However, due to extinction through electron scattering in the wind, the influence of a particular limb darkening law on the results is negligible. This is confirmed by our numerical tests with no limb darkening. Albedos were set to unity as appropriate for radiative atmospheres. We set the wind velocity parameter $\beta=1$ as in the L96 model. $\lambda$ was set to the central wavelength of the {\em BRITE} satellite red pass-band (6200 \emph{\AA}).

Usually the filling factors $\mu$ are of greatest interest in binary studies along with the inclination angle, masses and temperatures. However, the obtaining of reliable estimates for the values of these parameters is hampered by the fact that even in systems with two eclipses, they are correlated with one another, most notably the inclination. In addition, in the A13 model, there is an additional parameter degeneracy due to the wind. Given that WR22 has only one shallow eclipse, we therefore had to fix the filling factors by setting fixed radii of both components.

As with the L96 model, we adopted the value from \cite{Hamann19} for the radius of the WR component. In that paper, the authors did not fit each studied WR star individually, and instead pre-computed a grid of stellar atmosphere models against which they visually compared the observed WR spectra to find the closest grid model. They state that the uncertainty in the selection of the final model is plus or minus one grid mesh (the grid step size is 0.1\,dex). Clearly, the uncertainty defined in this manner is very informal. In any case, it represents the internal uncertainty of the model, and inaccuracies in observed parameters (e.g. distance) may further influence the final uncertainty of the model parameters. However, given the informal nature of the internal uncertainty, it is very hard to determine how it would be affected by such additional sources of uncertainty. For this reason, we will consider the informal uncertainty given in \cite{Hamann19} as a very rough estimate of the WR radius accuracy.

Since the A13 model does not assume a point-source O star, we also have to adopt a radius for that component of the system. This radius depends sensitively on the O-star luminosity class (V or III). \cite{Schwei99} noted that the ratios of equivalent widths of several spectral lines hinted towards luminosity class III. On the other hand, by assuming that the eclipse in WR22 was total (the exact shape of the eclipse was unknown at the time), they obtained a flux ratio which pointed towards luminosity class V. They favoured the latter based on the marginal quality of their equivalent-width measurements. However, from the absence of a flat bottom in the eclipse shape of the present data, it is reasonable to assume that the eclipse is not total, although the eclipse could be total and not appear as such if the time spent as total is no longer than a few $\sim 100-min$ {\em BRITE} orbits and if during that time the stochastic WR variability due electron scattering off clumps in the winds masks its photometric signature.

In view of the uncertainty in the luminosity class of the O star, we ran our fitting procedure for two values of $R_{\rm O}$ corresponding to luminosity class V or III (Models 1 and 2, respectively). The radii were assigned to the respective mean deduced radii of large samples of O9 stars by \cite{Martins05}. The list of major model parameters (assumed, fitted, and derived) is presented in Table~\ref{tab:RocheWindpars}. As in the L96 model, a single uncertainty of $\sigma=0.0105$ was assigned to all data points.

The principal difference between the two models is the finite size of the O star in the A13 model, as opposed to the point-source O star approximation in the L96 model. Thus, in the A13 model, the eclipse may be made up of two components : (1) a partial geometric eclipse and (2) the decrease of the O star flux due to electron scattering in the wind. Also, the flux ratio $F_r$ cannot be set as an input parameter, since the O-star flux is affected by heating caused by the radiation of the more luminous WR star. Thus, the O/WR flux ratio can be computed only after a fit is done. For this reason, we compute a grid of models similarly to what was done with the L96 model, but for a range of $T_{\rm WR}$ instead of $F_r$.

One of the main difficulties in modeling light curves of eclipsing binaries is the accurate determination of the stellar temperatures. A common practice is to assume the temperature of one star based e.g. on its spectrum \citep{kall09}. We therefore fixed the better-known temperature of the O star according to an average O9V and O9III star for Models 1 and 2, respectively, from \cite{Martins05}.

A note on the WR temperature is appropriate here. Although it is said in item (viii) above that the temperature is a flux scaling parameter, it turns out that the spectrum of the WN7h star in WR22 does not differ much from the spectrum of a black body in the UV and visual domains. We verified this by plotting the best-fit model spectrum of WR22 from \cite{Hamann19} (the spectral data were obtained from the website of the Potsdam group \citep{todt15}; the model for WR22 is MW WNL-H50 06-08) versus a black body spectrum of a star with the radius and temperature equal to those from the above paper. The two spectral shapes are very similar except in the far IR range. The flux of the black body spectrum in the {\em BRITE} band $5500 - 7000$\AA~ is 0.935 of the Hamann et al. model spectrum. Thus, despite using the black body spectrum to calculate the WR monochromatic flux, the WR temperature of our best-fit model should not be very different from the temperature determined from the spectrum modeling $T_{\rm WR}=44\,700$\,K \citep{Hamann19}.

\begin{table}
\caption{Major A13 model parameters.}
\label{tab:RocheWindpars}

\begin{threeparttable}
\centering

\begin{tabular}{c c} 
Parameter & Value \\
\hline
\multicolumn{2}{c}{Assumed Roche model parameters} \\
\hline
$P$ [days] & 80.336 \tnote{a} \\
$T_0$ [HJD] & 2 450 127.47 \tnote{a} \\
$M_{\rm O}\sin^3i$ [$\rm M_\odot$] & $20.6$ \tnote{a} \\
$M_{\rm WR}\sin^3i$ [$\rm M_\odot$] & $55.3$ \tnote{a}  \\
e & 0.598 \tnote{a} \\
$\omega_{\rm O}$ [$^{\circ}$] & 88.2 \tnote{a}  \\
$T_{\rm O}$ [$\rm K$] & 32\,900 (Model 1); 31\,850 (Model 2) \tnote{b} \\
$T_{\rm WR}$ [$\rm K$] & 20\,000 - 160\,000\ \\
$R_{\rm O}$ [$\rm R_\odot$] & $7.53$ (Model 1); $13.38$ (Model 2) \tnote{b} \\
$R_{\rm WR}$ [$\rm R_\odot$] & $22.65$ \tnote{c} \\
\hline
\multicolumn{2}{c}{Assumed wind parameters} \\
\hline
$\beta$ & 1.0 \\
$V_{\infty}$ [$\rm km\, s^{-1}$] & 1785 \tnote{c} \\
$\mu_e$ & 1.39 \tnote{c} \\
\hline
\multicolumn{2}{c}{Best-fit parameters, Model 1 (WR + O9V)} \\
\hline
$T_{\rm WR}$ [$\rm K$] & $50\,000\pm 600$ \tnote{d} \\
$i$ [$^{\circ}$] & $83.5 \pm 0.4$ \\
$\tau_0$ & $0.0064 \pm 0.0006$ \\
$\Delta m_0$ & $-0.0007 \pm 0.0003$ \\
$\Delta\phi$ & $-0.0080 \pm 0.0004$ \\
$\chi^2/{\rm d.o.f}$ & $3830.95/3442 = 1.113$ \\
\hline
\multicolumn{2}{c}{Derived parameters, Model 1 (WR + O9V)} \\
\hline
$F_{\rm O}/F_{\rm WR}$ & $0.064\pm 0.002$ \tnote{d}\\
$\dot M_{\rm WR}$ [$10^{-5}\rm M_\odot$/year] & $1.86 \pm 0.2$ \\
$\mu_{\rm O}$ & $0.184$ \\
$\mu_{\rm WR}$ & $0.340$ \\
$M_{\rm O}$ [$\rm M_\odot$] &  $21.00$ \\
$M_{\rm WR}$ [$\rm M_\odot$] &  $56.38$ \\
$\log(g)_{\rm O}$ \tnote{e} & $4.01$ \\
$\log(g)_{\rm WR}$ \tnote{e} & $3.05$ \\
\hline
\multicolumn{2}{c}{Best-fit parameters, Model 2 (WR + O9III)} \\
\hline
$T_{\rm WR}$ [$\rm K$] & $100\,000\pm 1500$ \tnote{d} \\
$i$ [$^{\circ}$] & $78.8 \pm 0.4$ \\
$\tau_0$ & $0.0191 \pm 0.0007$ \\
$\Delta m_0$ & $0.00007 \pm 0.00002$ \\
$\Delta\phi$ & $-0.0080 \pm 0.0004$ \\
$\chi^2/{\rm d.o.f}$ & $3763.61/3442 = 1.087$ \\
\hline
\multicolumn{2}{c}{Derived parameters, Model 2 (WR + O9III)} \\
\hline
$F_{\rm O}/F_{\rm WR}$ & $0.085\pm 0.002$ \tnote{d}\\
$\dot M_{\rm WR}$ [$10^{-5}\rm M_\odot$/year] & $5.6 \pm 0.2$ \\
$\mu_{\rm O}$ & $0.323$ \\
$\mu_{\rm WR}$ & $0.336$ \\
$M_{\rm O}$ [$\rm M_\odot$] & $21.82$ \\
$M_{\rm WR}$ [$\rm M_\odot$] & $58.57$ \\
$\log(g)_{\rm O}$ \tnote{e} & $3.52$ \\
$\log(g)_{\rm WR}$ \tnote{e} & $3.07$ \\
\hline
\end{tabular}

\begin{tablenotes}
\item[a] \cite{Schwei99}
\item[b] \cite{Martins05}
\item[c] \cite{Hamann19}
\item[d] $1-\sigma$ error was estimated by exploring $\chi2$ in the vicinity of the best-fit grid point.
\item[e] Logarithm of the surface gravity.

\end{tablenotes}

\end{threeparttable}

\end{table}

\begin{table}
\caption{Best-fit A13 Model 1 parameters for three individual eclipses.}
\label{tab:RocheWindpars_ecl}

\begin{threeparttable}
\centering

\begin{tabular}{c c c c} 
Eclipses fitted & 1 & 2 & 3 \\
\hline
Parameter & \multicolumn{3}{c}{Value} \\
\hline
\multicolumn{4}{c}{Best-fit parameters, Model 1 (WR + O9V)} \\
\hline
$T_{\rm WR}$ [$\rm K$] & $52\,500\pm 1100$  \tnote{a} & $42500\pm 800$ \tnote{a} & $45000\pm 1300$ \tnote{a} \\
$i$ [$^{\circ}$] & $83.5 \pm 0.4$ & $83.1 \pm 0.6$ & $83.5 \pm 0.6$ \\
$\tau_0$ & $0.0049 \pm 0.0008$ & $0.0054 \pm 0.0010$ & $0.0059 \pm 0.0009$ \\
$\Delta m_0$ & $-0.001 \pm 0.001$ & $-0.0007 \pm 0.0003$ & $-0.0000 \pm 0.0005$ \\
$\Delta\phi$ & $0.0060 \pm 0.0006$ & $0.0082 \pm 0.0005$ & $0.0080 \pm 0.0005$ \\
$\chi^2/{\rm d.o.f}$ & $914.48/884=$ & $ 931.63/887=$ & $ 2029.91/1755=$ \\
 & $=1.034$ & $=1.050$& $=1.156$ \\
\hline
\multicolumn{4}{c}{Derived parameters, Model 1 (WR + O9V)} \\
\hline
$F_{\rm O}/F_{\rm WR}$ & $0.060\pm 0.003$ & $0.070\pm 0.002$ & $0.073\pm 0.003$ \\
$\dot M_{\rm WR}$ & $1.4 \pm 0.2$ & $1.5 \pm 0.3$ & $1.7 \pm 0.2$ \\
~[$10^{-5}\rm M_\odot$/year] & \multicolumn{3}{c}{ } \\
$\mu_{\rm O}$ & $0.184$ & $0.181$ & $0.184$ \\
$\mu_{\rm WR}$ & $0.340$ & $0.334$ & $0.340$ \\
$M_{\rm O}$ [$\rm M_\odot$] & $21.00$ & $21.05$ & $21.00$ \\
$M_{\rm WR}$ [$\rm M_\odot$] & $56.38$ &  $56.51$ &  $56.38$ \\
$\log(g)_{\rm O}$ \tnote{b} & $4.01$ & $4.01$ & $4.00$ \\
$\log(g)_{\rm WR}$ \tnote{b} & $3.05$ & $3.05$ & $3.05$ \\
\hline
\end{tabular}

\begin{tablenotes}
\item[a] $1-\sigma$ error was estimated by exploring $\chi2$ in the vicinity of the best-fit grid point.
\item[b] Logarithm of the surface gravity.

\end{tablenotes}

\end{threeparttable}

\end{table}

%[$10^{-5}\rm M_\odot$/year] & \multicolumn{2}{c}{ } \\

\begin{figure*}
\includegraphics[width=\textwidth]{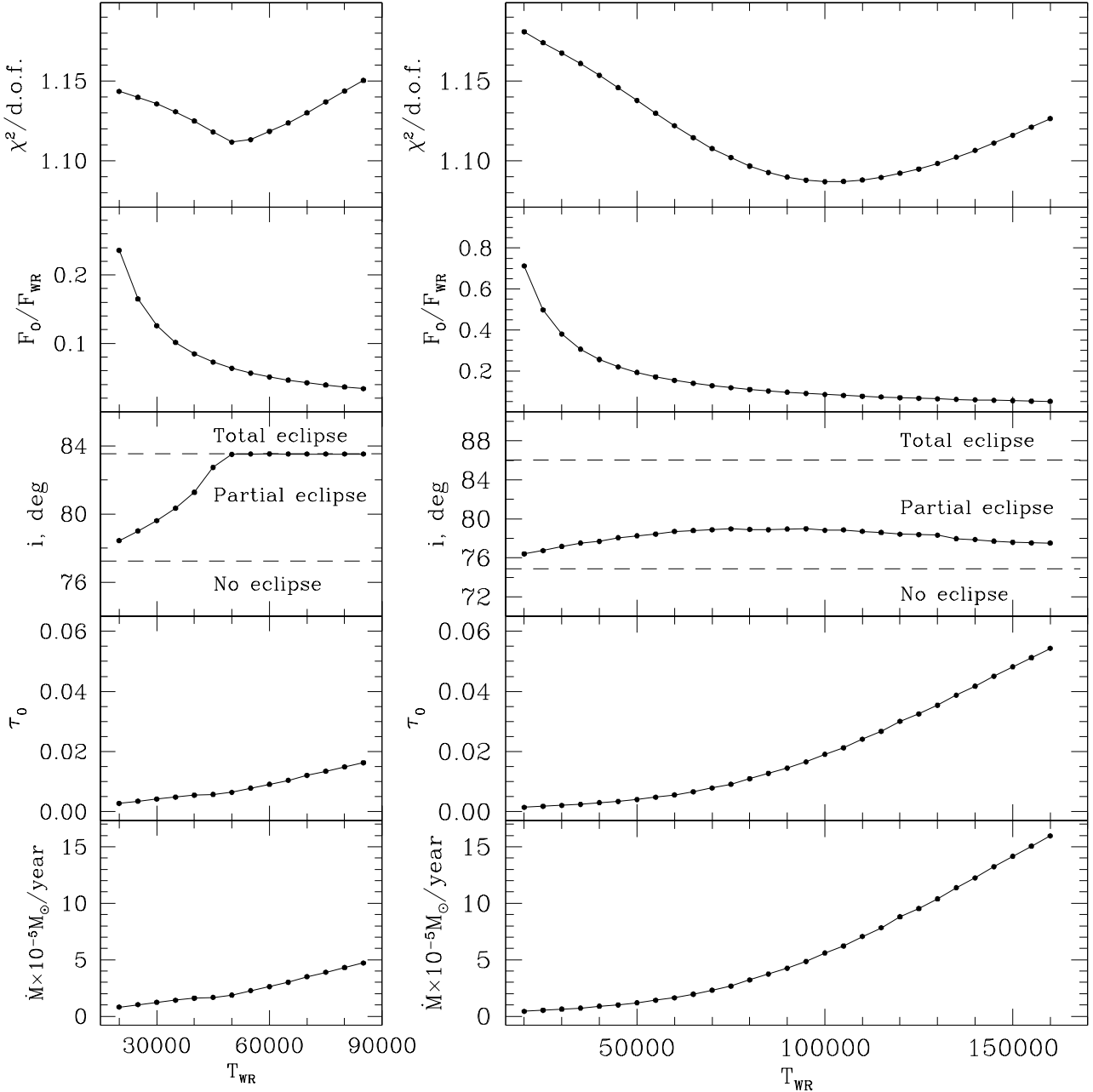}

\caption{Best-fit model parameters as a function of the WR temperature. Left: Model 1, $R_{\rm O} = 7.53 \rm R_\odot$ (luminosity class V). Right: Model 2, $R_{\rm O} = 13.38 \rm R_\odot$ (luminosity class III). The dashed lines in the inclination plots mark the borders between wind-only, partial geometric, and total eclipses.}
\label{fig:roche_pars}
\end{figure*}

\begin{figure*}
\includegraphics[width=\columnwidth]{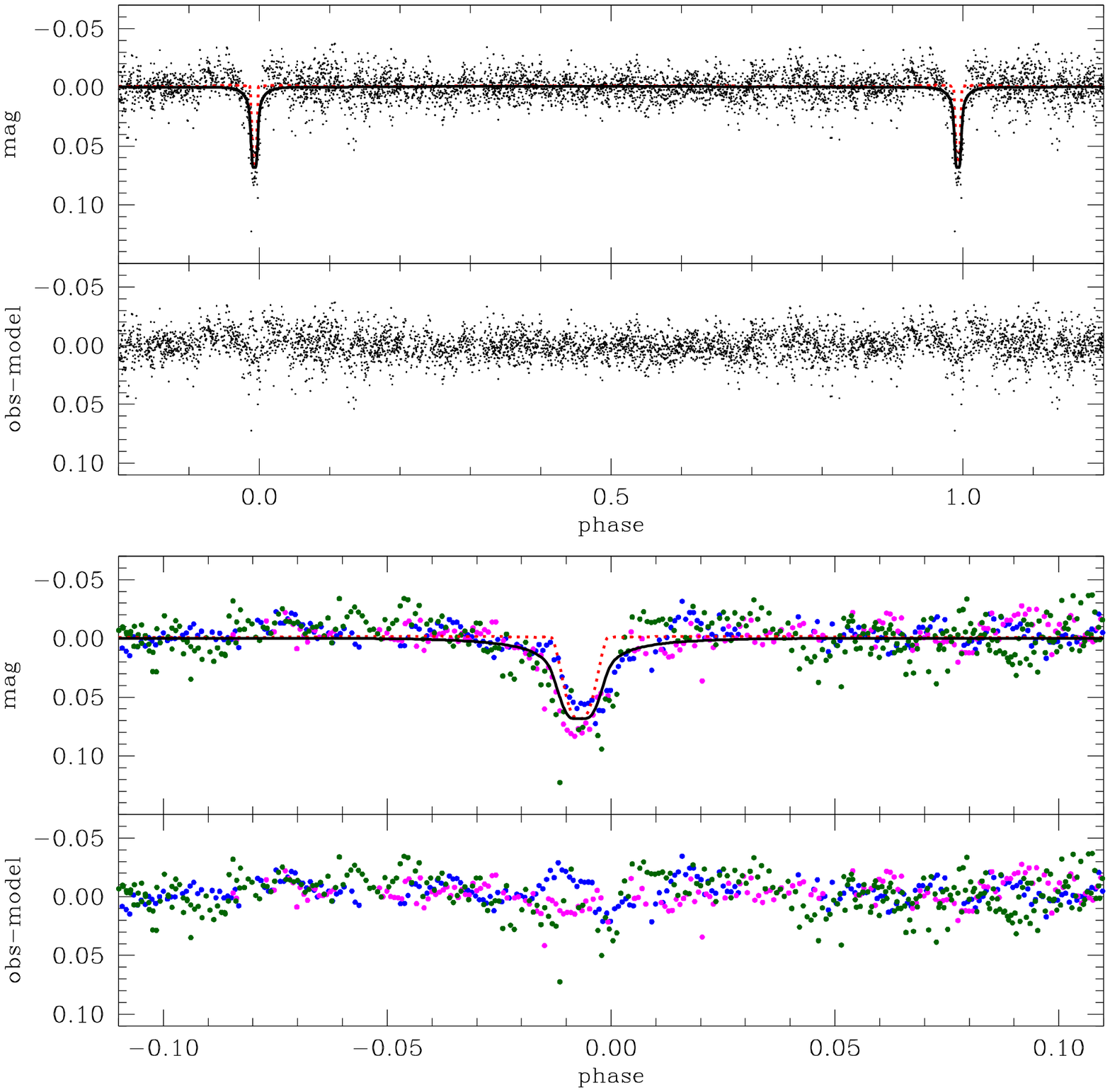}
\includegraphics[width=\columnwidth]{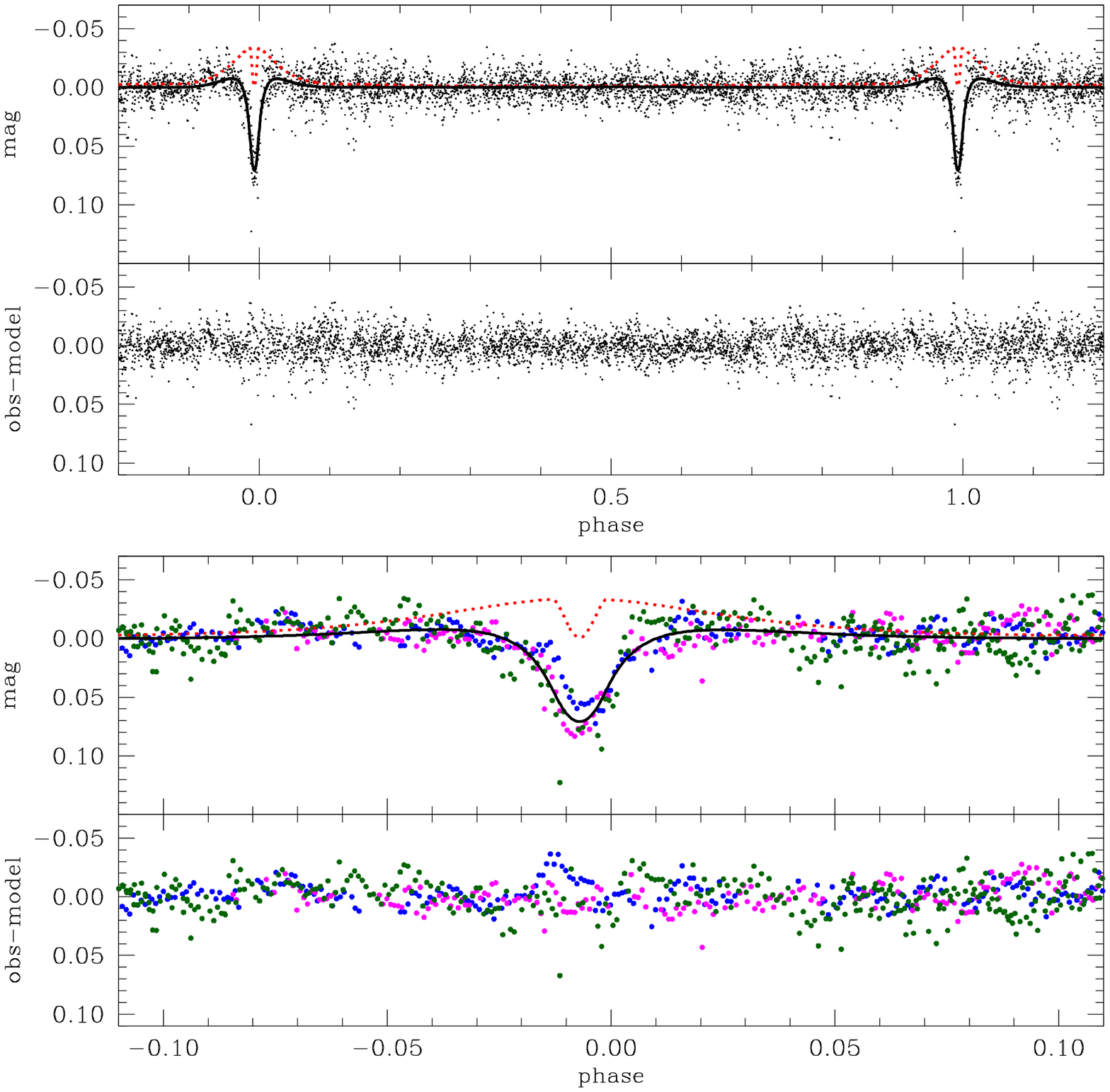}

\caption{Best-fit light curves corresponding to the respective $\chi^2$ minima from the previous figure. Left:  Model 1. Right: Model 2. The black solid line is the best-fit Roche+Wind light curve. The red dotted line shows the Roche model component only, i.e. not including the decrease of the O star flux due to scattering in the WR wind. The colours of the observed data points in the lower panel correspond to those in the L96 best-fit plot in Fig.\ref{Figure 7}.} 
\label{fig:roche_bestfit}
\end{figure*}

\begin{figure*}
\includegraphics[width=\textwidth]{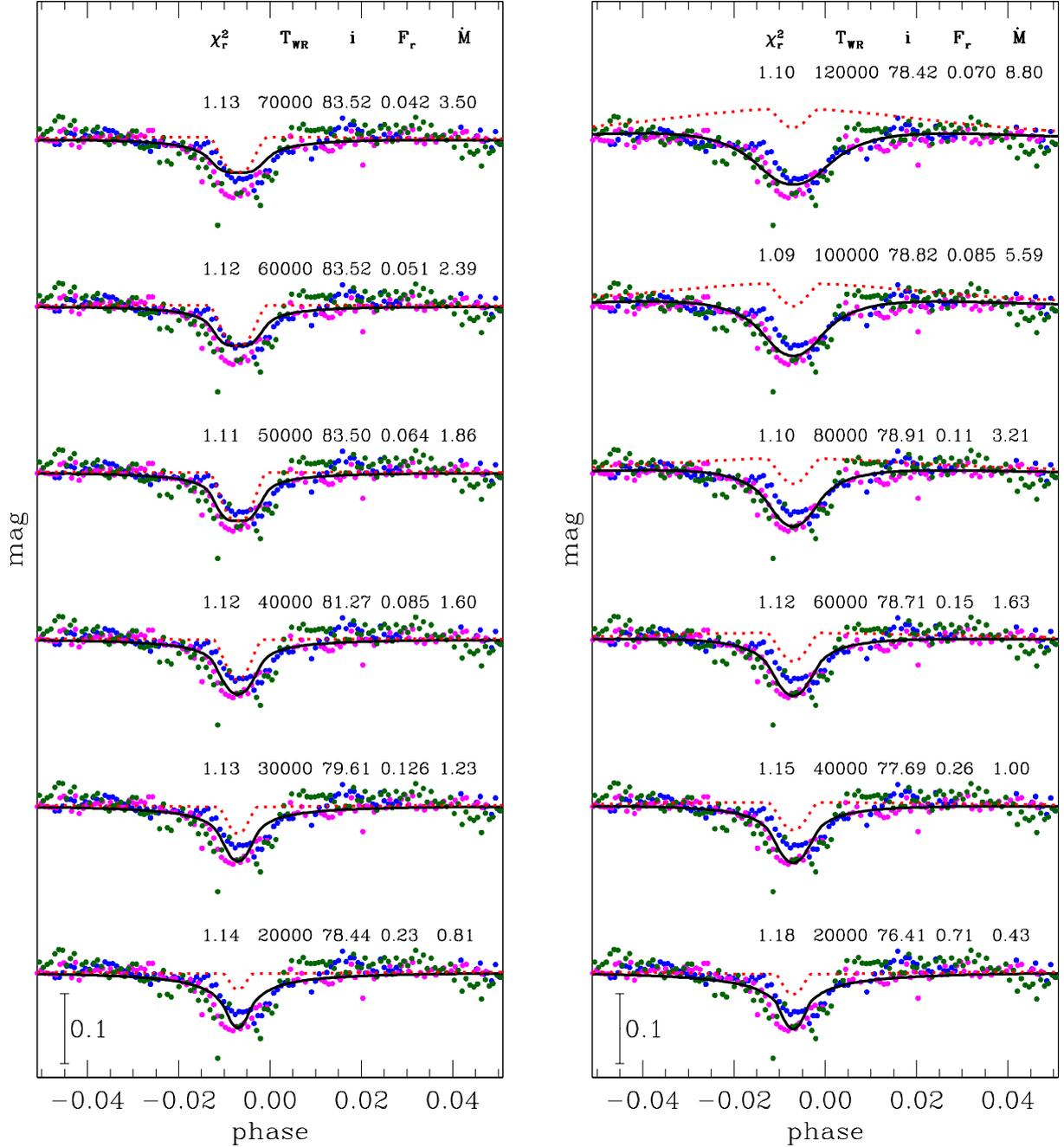}
\caption{Characteristic best-fit light curves at different $T_{\rm WR}$ for Model 1 (left) and 2 (right). The observed light curve is shown by coloured dots (the colour legend as above). The model light curves are shown by solid black lines. The red dotted lines show the Roche component of the light curves, i.e. not including the decrease of the O star flux due to scattering in the WR wind. The values of reduced $\chi^2$, $T_{\rm WR}$, inclination, flux ratio, and $\dot{M}$ (in units $10^{-5}$\,$\rm M_\odot$/year), respectively, are shown for each individual light curve.}
\label{fig:roche_lcgrid}
\end{figure*}

For each tested $T_{\rm WR}$, the best-fit solution for the light curve was found by using the Levenberg-Marquardt \citep{LM16} method. The covariation matrix was computed to estimate the errors of the best-fit parameters. As no total eclipse is observed, we used a penalty function restricting the orbital inclination by the maximal value defined by the total eclipse condition $r\cos(i_{\rm max}) = R_{\rm WR}-R_{\rm O}$, where $r$ is the orbital separation at the superior conjunction of the O star. The resulting fits of Models 1 and 2 for the chosen grid of WR temperatures are shown in Figs. \ref{fig:roche_pars}-\ref{fig:roche_lcgrid}, and the best-fit and derived parameters are presented in Tables~\ref{tab:RocheWindpars} and \ref{tab:RocheWindpars_ecl}.  Note that the value of reduced $\chi^2$ in Fig.~\ref{fig:roche_pars} changes just by a few hundredths between the best and worst fit models. However, the change is still significant, since at a large number of degrees of freedom (3442 in our case), even a change of reduced $\chi^2$ by 0.01 results in a large change of the null hypothesis probability, i.e. of the significance level of a model.

\begin{figure}
\includegraphics[width=\columnwidth]{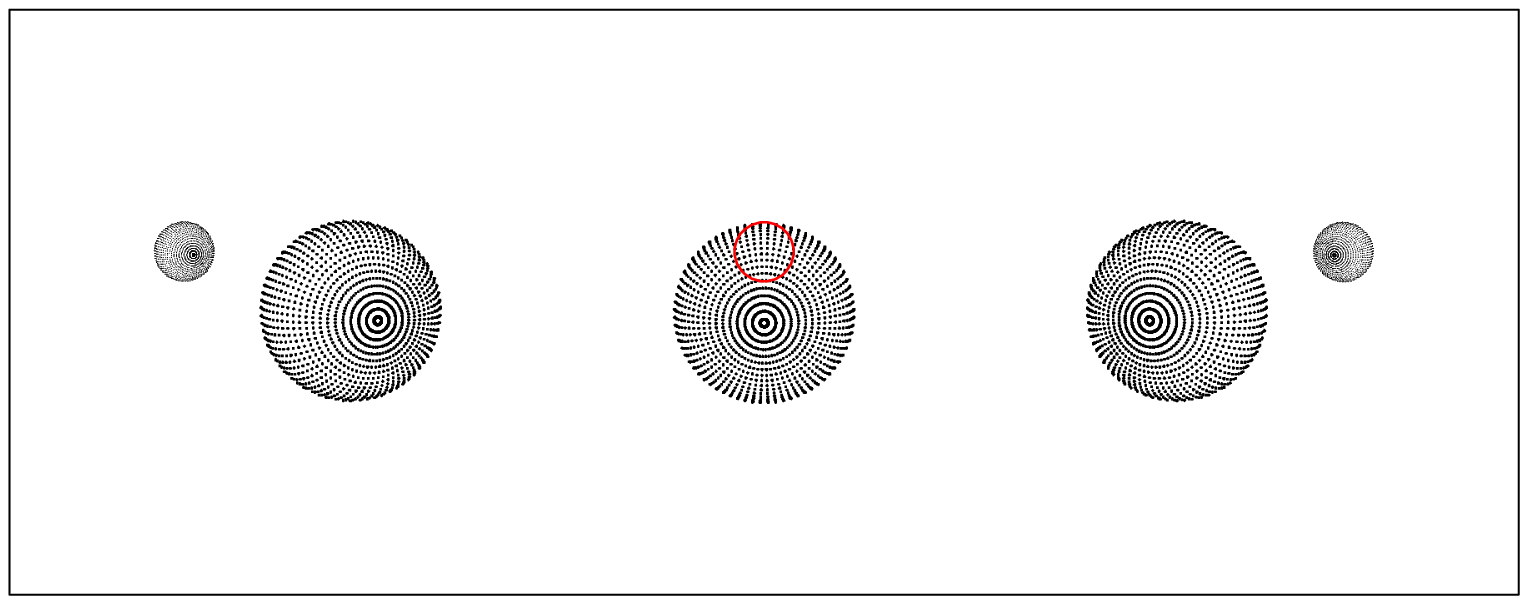}
\includegraphics[width=\columnwidth]{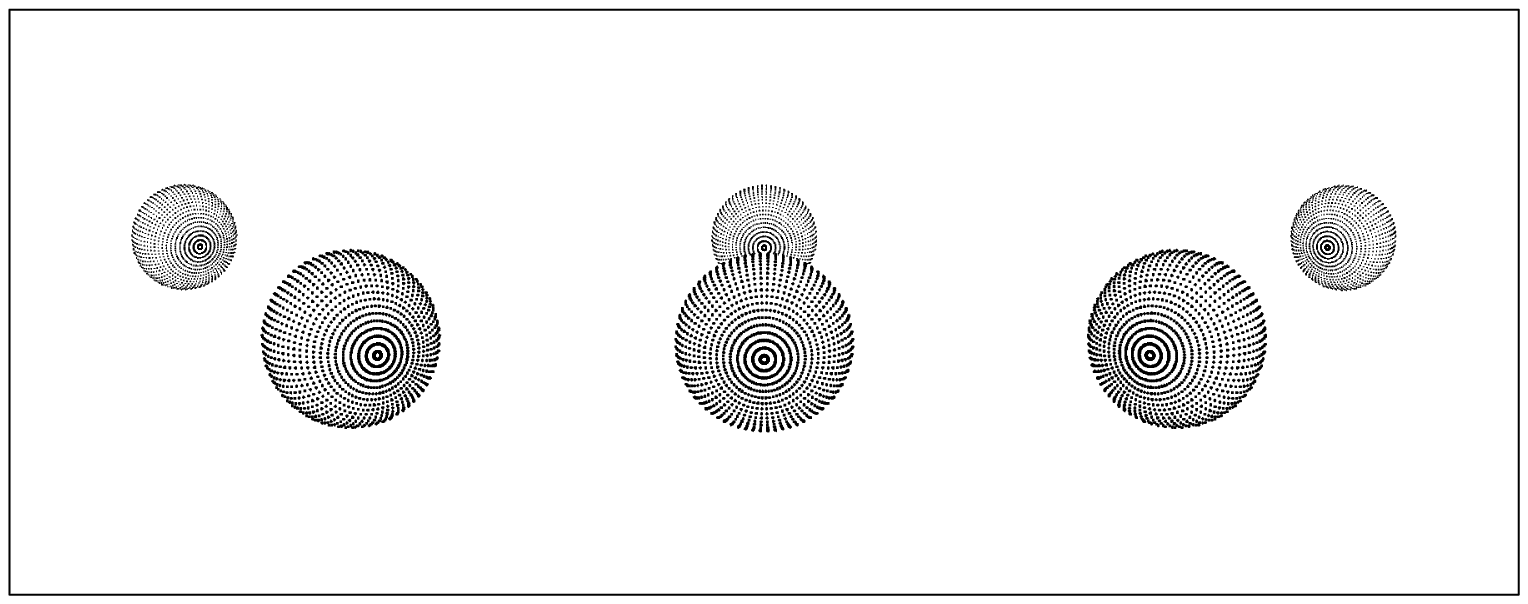}

\caption{Sky plane view of the system for best-fit A13 models at orbital phases before, at the moment of, and after conjunction. Top panel: Model 1 (WR + O9V), bottom panel: Model 2 (WR + O9III). The WR star is in front. The phase difference of the side plots with the conjunction moment (central plot) is $\pm 0.01$. The picture does not include the WR wind.}
\label{fig:roche_skyview}
\end{figure}

The behaviour of Models 1 and 2 are somewhat different. We consider them in turn:

{\bf Model 1}. At low WR temperatures, the best fit is achieved by the combination of a partial geometric eclipse and wind scattering of flux from the non-eclipsed parts of the O star disk. The reason is that at smaller inclination, atmospheric-only scattering would make the model eclipse wider than observed, while at larger inclinations the eclipse would be too deep because of the high flux ratio $F_{\rm O}/F_{\rm WR}$. As the WR temperature (and hence its flux) increases, the flux ratio $F_{\rm O}/F_{\rm WR}$ decreases and one needs a larger geometric eclipse (hence larger inclination). As the non-eclipsed part of the O-star disk becomes smaller, one needs to increase the optical depth of the wind to fit the eclipse ingress and egress. The overall best fit is achieved at $T_{\rm WR}=50\,000$\,K, where the flux ratio $F_{\rm O}/F_{\rm WR} = 0.064$ (see Table~\ref{tab:RocheWindpars}). The inclination angle is just marginally smaller than the critical angle where the total eclipse begins (the critical angle is defined by the condition that the whole O star disk is eclipsed and touches the circumference of the WR disk at the moment of conjunction: for Model 1 it is equal to $83.53^\circ$). At even higher WR temperatures, the flux ratio $F_{\rm O}/F_{\rm WR}$ becomes so small that even a total eclipse is unable to reproduce the eclipse depth. This is why at these WR temperatures the inclination angle remains constant and equal to the critical value.

{\bf Model 2}. At low WR temperatures, the behavior of this model is qualitatively similar to Model 1. For the same reasons as with the latter, the best fit is achieved when the model eclipse is a combination of wind scattering and a geometric eclipse. The inclination angle is smaller than that found in Model 1 for the same WR temperatures, since the O star radius is about 1.7 times larger (thus requiring a smaller inclination for the geometric overlapping of the stellar disks). By increasing the WR temperature, the inclination angle increases more slowly than in Model 1, again because of the larger O star radius, so smaller inclination angles are needed to reach the same eclipse depth. The flux ratio $F_{\rm O}/F_{\rm WR}$ in Model 2 at a given WR temperature is roughly proportional to $(R_{\rm O}/R_{\rm WR})^2$, and so is higher than in Model 1. For this reason, the combination of partial geometric eclipse and wind scattering remains relevant at higher WR temperatures than in Model 1. By increasing the WR temperature, the reflection effect becomes more and more important (see Fig.~\ref{fig:roche_lcgrid}). To compensate for this effect, the wind optical depth has to be increased, leading to an increase of the eclipse depth, which in turn is compensated by a slow decrease of the inclination angle starting from $T_{\rm WR}\sim 80\,000$\,K (see Fig. \ref{fig:roche_pars}, right panel). The overall best fit is achieved at $T_{\rm WR}=100\,000$\,K, where the flux ratio $F_{\rm O}/F_{\rm WR} = 0.085$ (see Table~\ref{tab:RocheWindpars})

In Fig. \ref{fig:roche_skyview} the sky plane view of the overall best-fit Models 1 and 2 from Table~\ref{tab:RocheWindpars} is shown at orbital phases immediately before, at the moment of, and after superior conjunction of the O star.

The three observed eclipses appear to be somewhat different from one another when taken at face value. We explain in the next section however that the differences in the intrinsic eclipse shapes are only apparent, being the result of stochastic variability caused by clumping in the WR wind, as seen outside the eclipses, where one sees variations with similar amplitude and timescale. To check the sensitivity of the A13 model parameters when applied to individual eclipses, we carried out a fitting procedure similar to that previously described for Model 1 for each of the three orbital cycles containing the observed eclipses. As before, all observed data points were assigned the empirical $\sigma$ calculated between phases 0.1 and 0.9 of the corresponding orbital cycle. The empirical standard deviations for the first, second, and third orbital cycles are 0.0088, 0.0099, and 0.0114 respectively. The results for the best overall fits are shown in Table~\ref{tab:RocheWindpars_ecl}. As expected, in the case of the first orbital cycle, the overall best fit is achieved at a lower flux ratio and, consequently, a higher WR temperature. The reason for this is that the first eclipse appears to be shallower than the next two. This means that the  relative modeled contribution of the WR flux to the total system flux appears to be larger, resulting in the higher WR temperature. The second and third eclipses are deeper than the first one so the situation for them appears to be the opposite. The WR temperature of the solution for the complete data set is between that for the individual orbital cycles.

It may seem surprising that the WR mass-loss rates in the individual solutions are systematically smaller than the mass-loss rate of the combined solution. This result can be understood by observing the difference between the complete model light curves (shown in Fig.~\ref{fig:roche_lcgrid} by black solid lines) and the model light curves without any wind contribution (shown in the same figure by red dotted lines). It is the difference between these curves which defines the wind optical depth. The individual eclipses are slightly shifted in phase (see $\Delta\phi$ parameter in Table~\ref{tab:RocheWindpars_ecl}) and have different depths. As a result, the best-fit model eclipse of the combined data is wider that the best-fit model eclipses of each individual data set. Yet the geometrical eclipse remains basically the same, as the inclination angles of all solutions are very similar. Thus the difference between the ``no wind'' and ``wind'' model light curves is larger for the complete data set, resulting in larger $\tau_0$ and the mass-loss rate.

\section{Discussion}
\label{section:Discussion}

As stated in Section 3.3, the black body spectrum used by the best-fit A13 model when calculating the WR flux should have a temperature close to the value found by spectral modeling of $T_{\rm WR} = 44\,700$\,K \citep{Hamann19}. This means that although the best fit in Model 2 is formally slightly better than in Model 1, Model 2 should be rejected because of the strong discrepancy between the model WR temperature and that of \cite{Hamann19}. In other words, our analysis shows that O star in WR22 has a luminosity class of V, not III. On the other hand, the WR temperature in Model 1 is quite close to that of \cite{Hamann19}. Therefore, we take the Model 1 results as the final light curve solution. It should be noted that while we made our choice between the models based on the WR temperature of \cite{Hamann19}, the latter depends on the brightness of the system. Thus we effectively made our choice based on the system brightness. A WR star with the adopted radius and a temperature of e.g. $10^5$\,K would simply be too bright.

Our best-fit model temperature of $T_{\rm WR}=50000$\,K is still larger than $T_{\rm WR}=44700$\,K from \cite{Hamann19}. This cannot be explained by the spacing in the grid of temperatures used in our model. At $T_{\rm WR}=45000$\,K, close to the \citeauthor{Hamann19} value, the value of our reduced $\chi^2$ is larger. The ratio of our model WR flux to that of \citeauthor{Hamann19} is about $(50000/44700)^4\sim 1.5$. This may mean that our model WR temperature is overestimated, which is possible if the assumed O9V star temperature \citep{Martins05} is overestimated as well. To bring our model WR temperature to $44700$\,K, we would have to reduce the O9V star temperature to $29400$\,K. This value is too low considering the accuracy of stellar temperatures in \cite{Martins05}. Another explanation may be related to variations of the eclipse shape and depth from one orbital cycle to another (see below). Thus, the three observed eclipses might not reveal the regular variability at the eclipse phases with sufficient accuracy. Another possibility is that either the WR radius of the \citeauthor{Hamann19} model or its temperature are underestimated. Recall that the model of \citeauthor{Hamann19} does not account for the O9V (lower temperature) star contribution to the total flux of the system. However, since the flux ratio is small, this explanation is not very likely.

Note that in our Model 1 $\dot{M}_{\rm WR}=1.86\cdot 10^{-5}\rm M_\odot/year$, which is approximately two times less than $\dot{M}_{\rm WR}=3.98\cdot 10^{-5}\rm M_\odot/year$ in \cite{Hamann19}. The mass loss rate in spectral modeling depends on the adopted clumping model and its magnitude. The clumping model used most commonly is a simple model of optically thin clumps and a constant clumping contrast $D$ (inverse to the spatial filling factor). In this case, the modeling of spectral lines actually determines the value of $\dot{M}\sqrt{D}$. In \cite{Hamann19}, $D = 4$. As noted by the authors \citep{Hamann06}, this value is rather conservative: ``There are indications that the clumping is actually even stronger, and hence the mass loss rates might still be overestimated generally by a factor of 2 or 3''. \cite{hillier20} also notes that values of $D=10$ or $20$ are ``routinely used in the literature'' (see the above paper for more details on difficulties of determining mass loss rates from spectral modeling). Clumping mainly affects spectral lines, as the recombination rate is proportional to the square of density. As electron scattering has a linear dependence on density, our derived mass-loss rate is insensitive to clumping.

The flux ratio of our best-fit Model 1 is $0.064$ (Table~\ref{tab:RocheWindpars}). \cite{Rauw96} obtained a flux ratio of $\sim 0.08-0.2$ (at the reference wavelength 5500\AA) from different spectral lines. Taking into account the accuracy of measurements, they give the mean luminosity ratio $F_{\rm WR}/F_{\rm O} = 8.2$ (corresponding to $F_{\rm O}/F_{\rm WR} = 0.122$).  \cite{Schwei99} estimated the lower limit of the flux ratio as being $\sim 0.08$, assuming that the eclipse is total. Thus, our estimate of the flux ratio seems to be smaller than those above. However, the estimate of \cite{Schwei99} was based on the adopted eclipse depth of 0.083 taken from the {\em b} (4670\AA) light curve of \cite{Gosset91}, adjusted to {\em y} (5470\AA) by using $\Delta(b-y)$. The eclipse in the original light curve is rather poorly defined, so its actual depth is hard to estimate accurately. Our eclipses have a smaller depth (especially the first one), thus the estimate of \cite{Schwei99} should probably be reduced. These authors also note that in their spectra the O-star absorption lines are weaker than those in \cite{Rauw96}, i.e. for the He\,I 4471 line the equivalent widths are 0.057\AA~and 0.035\AA, for He\,II 4542  0.052\AA~and 0.016\AA~respectively. This means that the flux ratio of \cite{Rauw96} might be overestimated. Note also that to estimate the flux ratio, the latter authors had to account for the different slopes of O and WR star spectra and used a ``typical'' equivalent width of a single O-star`s spectral lines (which they did by using a power-law approximation of the WR spectrum and Kurucz's model for the O-star spectrum) and used average equivalent widths of spectral lines taken from a list of lines of single O-stars shown in \cite{Conti73}. The uncertainties involved in the procedure make it hard to reliably estimate the actual error of the flux ratio. Note also that the reference wavelengths of both \cite{Rauw96} and \cite{Schwei99} are smaller than ours, although this probably has only a minor effect. We conclude that the flux ratio obtained in the current study is at least not in strong contradiction with the previous estimates.

In A13 Models 1 and 2, we have used a single fixed WR radius and two fixed radii of the O star. Evidently, the actual stellar radii may be quite different from these values. How strongly will the change in radii affect the results? We already saw that increasing the O-star radius by $\sim 1.7$ times significantly changed the results. As we discussed in the previous section, the uncertainty of the WR radius can be very roughly and informally estimated as $\sim 0.1$\,dex. Our Models 1 and 2 correspond to two values of $R_{\rm WR}/R_{\rm O}$. If other WR and/or O star radii are suggested, one could roughly estimate the expected results by interpolating  between our Models 1 and 2, to the desired radius ratio.

We used the L96 model along with A13, since the former was used in numerous studies of \mbox{WR + OB} binaries with small orbital inclination angles. So it was interesting to see how well this model can handle a situation where the inclination is large so that a photospheric eclipse could occur. It turns out that the solution in the L96 atmospheric model gives an inclination angle equal to the critical value such that the projection of the O star point-source on the sky plane just touches the WR disk at superior O-star conjunction. With such a solution, it is clear that the assumption of the point-source O star is not appropriate. Still, it seems that the L96 model can be used to quickly estimate approximate parameters. It should be borne in mind that in the case of obtaining a critical inclination angle, its value should be increased by a few degrees. The fact that the L96 model produces a slightly worse fit in comparison with that of the A13 model is explained by the point-source approximation for the O star; this makes the model eclipse narrower than the observed one.

A visual inspection of the model and observational light curves in Figs.~\ref{Figure 7},\,\ref{fig:roche_bestfit} gives the impression that the scatter of the observational data points during the eclipse is larger than outside of it. This may be due to two reasons: an actual change of the eclipse shape over time, allowing the depth to increase from the first to the second observed eclipse, or an increased random scatter of data points during eclipses compared to parts of the light curve outside of eclipses. If the first reason is correct, it could mean that the WR mass loss rate has increased during the time elapsed from the first to the second eclipse. However, the required increase $\rm(1\div3)\times 10^{-6}\rm M_\odot$/year (Table~\ref{tab:RocheWindpars_ecl}) seems to be too large for a timescale of about a year. If the second is true, this may be due to the effect of scattering of the O-star emission by clumps in the WR wind. To verify this quantitatively, Table \ref{tab:ocecl} gives a comparison between the L96 and A13 Model 1 (the latter being applied to the full light curve and to the three eclipses separately) on the basis of scatter and reduced $\chi ^2$ for various key parts of the light curve. From this, we conclude the following:

\begin{enumerate}
    \item For the full light curve, A13 gives a better result, although not overwhelmingly so.
    \item During the part outside the eclipse (and excluding the wings) both L96 and A13 are equal (which is not surprising given that the observed flux is constant and both models produce a constant flux in this part).
    \item In the reflection wings and eclipse both A13 Model 1 and L96 computed for the whole light curve show large scatter (the former being somewhat smaller). The lack of systematic deviations in the residuals in either case suggests that the scatter is intrinsic to the clumpy nature of the WR wind.
    \item Solutions of the A13 Model 1 for the individual orbital cycles generally show large scatter in the eclipse and the wings. Two exceptions are: the eclipse of the first orbital cycle and the wings of the second cycle. However, the small (o-c) scatter in an eclipse is not supplemented by a small scatter in its wings and vice versa. The fact that the scatter in the eclipses and their wings is still generally large does not agree well with the assumption of changing mass-loss rate.
    \item Finally, we conclude that a likely explanation for the origin of an increased (o-c) scatter within the eclipse is O-star light scattered off WR-wind clumps during the eclipse, in contrast to WR starlight being scattered by WR-wind clumps during the whole orbit.
\end{enumerate}

The apparently variable data scatter in different parts of the light curve is the reason why we do not provide the significance levels of the models in our tables. All models are formally rejected at a significance level, for example, of 1 percent. However, this is apparently not due to a bad fit per se, but due to the fact that we apply the same data error to the entire light curve, while in reality it appears to be variable with orbital phase. To obtain a formally satisfactory fit, many more observations are required to provide an adequate estimate of the data scatter in different parts of the light curve. Then the scatter could be calculated as e.g. standard deviations in narrow phase intervals independently of any modeling. Until then, we have no reliable and model independent way of obtaining the data uncertainties at the eclipse phases of the light curve. Our analysis above of the (o-c) of individual eclipses implicitly assumed that A13 Model 1 was adequate and could be accepted. Then, the individual eclipses are statistically consistent, and the data scatter during the eclipses can be estimated by $\sigma\rm (o-c)$ of the best fit model of the complete observed light curve (A13 Model 1 in Table~\ref{tab:ocecl}).

Finally, we estimated a possible change in the period following from the discovered phase-shift, assuming that it is solely due to inaccuracy in the determination of the period. Given that the best-fit A13 Model 1 phase shift is equal to $-0.008P = -0.643$\,days, the period $80.336$\,days of \cite{Schwei99} should be decreased by $0.0065$\,days. Note that \cite{Schwei99} give the accuracy of the period $0.0013$\,days. However, they also give the error of $T_0=0.14$\,days, so it is difficult to say whether the observed phase shift is fully due to the inaccuracy of the period or if it is also in part due to an inaccuracy in the determination of $T_0$. Furthermore, the period of a WR+O binary should actually increase due to loss of angular momentum by the WR star through its stellar wind. This unknown potential period increase only adds to the uncertainty. For this reason, we did not attempt to refine the period value.

It should also be noted that the values of our best-fit model parameters are defined not only by the shape of the light curve, but also by the values of the assumed parameters. This is clearly demonstrated by the large difference between the WR temperatures in Models 1 and 2, even though their $1-\sigma$ errors are much smaller than the difference.

\begin{table}
\caption{Comparison of different parts of the WR22 light curve from the different models. L96 is the best-fit model from Table~\ref{Table 1}, A13 Model 1 from Table~\ref{tab:RocheWindpars} (all eclipses), A13 Model 1 (1)-(3) from Table~\ref{tab:RocheWindpars_ecl} (individual eclipses).}
%\scriptsize
\label{tab:ocecl}
\begin{threeparttable}
\centering
\begin{tabular}{c c c c} 
Orbit part   &  d.o.f.    &   $\sigma$(o-c)  & $\chi^2$/d.o.f. \\
\hline
   &     \multicolumn{3}{c}{L96}  \\
\hline
Full         & 3442 & 0.0114 & 1.17 \\
Outside ecl. & 2793 & 0.0105 & 1.01 \\
Eclipse      & 114  & 0.0186 & 3.15 \\
Wings        & 440  & 0.0134 & 1.62 \\
\hline
   &     \multicolumn{3}{c}{A13 Model 1} \\
\hline
Full         & 3442 & 0.0111 & 1.11 \\
Outside ecl. & 2793 & 0.0105 & 1.00 \\
Eclipse      & 114  & 0.0150 & 2.05 \\
Wings        & 440  & 0.0127 & 1.46 \\
\hline
  &     \multicolumn{3}{c}{A13 Model 1 (1)} \\
\hline
Full         & 884  & 0.0090 & 1.03 \\
Outside ecl. & 694  & 0.0089 & 1.00 \\
Eclipse      & 37   & 0.0077 & 0.76 \\
Wings        & 121  & 0.0101 & 1.31 \\
\hline
  &     \multicolumn{3}{c}{A13 Model 1 (2)} \\
\hline
Full         & 886  & 0.0102 & 1.05 \\
Outside ecl. & 718  & 0.0099 & 1.00 \\
Eclipse      & 37   & 0.0110 & 1.24 \\
Wings        & 109  & 0.0097 & 0.96 \\
  &     \multicolumn{3}{c}{A13 Model 1 (3)} \\
\hline
Full         & 1754 & 0.0123 & 1.16 \\
Outside ecl. & 1464 & 0.0114 & 1.00 \\
Eclipse      & 35   & 0.0210 & 3.39 \\
Wings        & 199  & 0.0157 & 1.89 \\
\hline
\end{tabular}
\begin{tablenotes}
\item The rms scatter used to calculate the $\chi^2$ values is based on the part of the light curve outside the eclipse (and wings). The part outside the eclipse/wings corresponds to \mbox{[0.1, 0.9]} for both the L96 and A13 models. The eclipse-phase interval corresponds to the calculated centre of the eclipse using either model $\pm$ 0.02. These are slightly different according to the particular fit. The wings are regions extending 0.07 in phase on each side of the eclipse for all models.
\end{tablenotes}
\end{threeparttable}
\end{table}

\section{Conclusions}

Thanks to the ability of the {\em BRITE} nano-satellites to lock onto a field for up to six months at a time, we were able to extract a complete, high-precision light curve for WR22 that reveals three complete single eclipses. The increased scatter in the eclipses is an interesting result and might be worth following up with other grazing eclipsers among WR+O systems. Considering the observed eclipse shape variations from one orbital cycle to another, it would be a worthy endeavor to acquire more eclipse observations of WR22. Unfortunately, while the length of an eclipse is moderately small (about three days), the long orbital period makes this task very time consuming as the accuracy of the mean light curve is proportional to the square root of the number of observations. Obtaining more spectra with high signal-to-noise ratio would also be helpful in improving the value of the flux ratio. These observations do not require especially lengthy efforts.

The more appropriate A13 photospheric plus wind model has shown that it is impossible to give a unique value of the orbital inclination and WR mass-loss rate based solely on photometric observations. However, the information on the WR temperature from the spectral analysis of \cite{Hamann19} allows us to make an unambiguous conclusion that the luminosity class of the O star is V, not III, and, therefore, to choose the unique solution (A13 Model 1). The best-fit orbital and stellar parameters obtained from this model, when fitted over the three {\em BRITE}-observed eclipses, are $i = 83.5 \pm 0.4^{\circ}$, $T_{\rm WR} = 50 000$ $K$ and $\dot M_{\rm WR} = (1.86 \pm 0.2) \times 10^{-5} \dot M_{\odot}/yr$.

%, which are all compatible with the values obtained from the spectral modeling of \cite{Hamann19}.

The L96 model, while being very simple and generally not suited for binaries with large inclination angles, can be used for quick and rough estimates of stellar and wind parameters.

\section*{Acknowledgements}

The authors are grateful to the anonymous referee for an insightful report, which helped us to significantly improve the results. IA and EA are grateful to Prof. S.A.Lamzin, Dr. A.V.Dodin, and Prof. W.-R.Hamann for helpful discussions. The work of IA (analysis with the A13 model) was supported by the RSF grant 17-12-01241 (Russia). The work of EA was supported by the Scientific Educational School of Lomonosov Moscow State University ``Fundamental and applied Space Research''. NSL \& AFJM are grateful to NSERC (Canada) for financial aid.
 
 \section*{Data Availability}
The de-trended {\em BRITE} data of WR22 underlying this article are available on Figshare, at https://dx.doi.org/10.6084/m9.figshare.14850135

%The Acknowledgements section is not numbered. Here you can thank helpful colleagues, acknowledge funding agencies, telescopes and facilities used etc. Try to keep it short.

%%%%%%%%%%%%%%%%%%%%%%%%%%%%%%%%%%%%%%%%%%%%%%%%%%



\bibliographystyle{mnras}
\bibliography{WR22} % if your bibtex file is called example.bib

\begin{thebibliography}{}
\makeatletter
\relax
\def\mn@urlcharsother{\let\do\@makeother \do\$\do\&\do\#\do\^\do\_\do\%\do\~}
\def\mn@doi{\begingroup\mn@urlcharsother \@ifnextchar [ {\mn@doi@}
  {\mn@doi@[]}}
\def\mn@doi@[#1]#2{\def\@tempa{#1}\ifx\@tempa\@empty \href
  {http://dx.doi.org/#2} {doi:#2}\else \href {http://dx.doi.org/#2} {#1}\fi
  \endgroup}
\def\mn@eprint#1#2{\mn@eprint@#1:#2::\@nil}
\def\mn@eprint@arXiv#1{\href {http://arxiv.org/abs/#1} {{\tt arXiv:#1}}}
\def\mn@eprint@dblp#1{\href {http://dblp.uni-trier.de/rec/bibtex/#1.xml}
  {dblp:#1}}
\def\mn@eprint@#1:#2:#3:#4\@nil{\def\@tempa {#1}\def\@tempb {#2}\def\@tempc
  {#3}\ifx \@tempc \@empty \let \@tempc \@tempb \let \@tempb \@tempa \fi \ifx
  \@tempb \@empty \def\@tempb {arXiv}\fi \@ifundefined
  {mn@eprint@\@tempb}{\@tempb:\@tempc}{\expandafter \expandafter \csname
  mn@eprint@\@tempb\endcsname \expandafter{\@tempc}}}

\bibitem[\protect\citeauthoryear{{Antokhin}}{{Antokhin}}{2012}]{ant12}
{Antokhin} I.~I.,  2012, \mn@doi [\mnras] {10.1111/j.1365-2966.2011.20054.x},
  \href {https://ui.adsabs.harvard.edu/abs/2012MNRAS.420..495A} {420, 495}

\bibitem[\protect\citeauthoryear{{Antokhin}}{{Antokhin}}{2016}]{ant16}
{Antokhin} I.~I.,  2016, \mn@doi [\mnras] {10.1093/mnras/stw2111}, \href
  {https://ui.adsabs.harvard.edu/abs/2016MNRAS.463.2079A} {463, 2079}

\bibitem[\protect\citeauthoryear{{Antokhina}}{{Antokhina}}{1988}]{ant88}
{Antokhina} E.~A.,  1988, \sovast, \href
  {https://ui.adsabs.harvard.edu/abs/1988SvA....32..608A} {32, 608}

\bibitem[\protect\citeauthoryear{{Antokhina}}{{Antokhina}}{1996}]{ant96}
{Antokhina} E.~A.,  1996, Astronomy Reports, \href
  {https://ui.adsabs.harvard.edu/abs/1996ARep...40..483A} {40, 483}

\bibitem[\protect\citeauthoryear{{Antokhina}, {Moffat}, {Antokhin}, {Bertrand}
  \& {Lamontagne}}{{Antokhina} et~al.}{2000}]{ant00}
{Antokhina} E.~A.,  {Moffat} A. F.~J.,  {Antokhin} I.~I.,  {Bertrand} J.-F.,
  {Lamontagne} R.,  2000, \mn@doi [\apj] {10.1086/308228}, \href
  {https://ui.adsabs.harvard.edu/abs/2000ApJ...529..463A} {529, 463}

\bibitem[\protect\citeauthoryear{{Antokhina}, {Antokhin}  \&
  {Cherepashchuk}}{{Antokhina} et~al.}{2013}]{ant13}
{Antokhina} E.~A.,  {Antokhin} I.~I.,   {Cherepashchuk} A.~M.,  2013,
  Astronomical and Astrophysical Transactions, \href
  {https://ui.adsabs.harvard.edu/abs/2013A&AT...28....3A} {28, 3}

\bibitem[\protect\citeauthoryear{{Balona}, {Egan}  \& {Marang}}{{Balona}
  et~al.}{1989}]{Balona89}
{Balona} L.~A.,  {Egan} J.,   {Marang} F.,  1989, \mn@doi [\mnras]
  {10.1093/mnras/240.1.103}, \href
  {https://ui.adsabs.harvard.edu/abs/1989MNRAS.240..103B} {240, 103}

\bibitem[\protect\citeauthoryear{{Cherepashchuk}, {Eaton}  \&
  {Khaliullin}}{{Cherepashchuk} et~al.}{1984}]{cher84}
{Cherepashchuk} A.~M.,  {Eaton} J.~A.,   {Khaliullin} K.~F.,  1984, \mn@doi
  [\apj] {10.1086/162156}, \href
  {https://ui.adsabs.harvard.edu/abs/1984ApJ...281..774C} {281, 774}

\bibitem[\protect\citeauthoryear{{Conti}}{{Conti}}{1973}]{Conti73}
{Conti} P.~S.,  1973, \mn@doi [\apj] {10.1086/151856}, \href
  {https://ui.adsabs.harvard.edu/abs/1973ApJ...179..161C} {179, 161}

\bibitem[\protect\citeauthoryear{{Conti}, {Niemela}  \& {Walborn}}{{Conti}
  et~al.}{1979}]{Conti79}
{Conti} P.~S.,  {Niemela} V.~S.,   {Walborn} N.~R.,  1979, \mn@doi [\apj]
  {10.1086/156837}, \href
  {https://ui.adsabs.harvard.edu/abs/1979ApJ...228..206C} {228, 206}

\bibitem[\protect\citeauthoryear{{Gamen} et~al.,}{{Gamen}
  et~al.}{2006}]{Gamen2006}
{Gamen} R.,  et~al., 2006, \mn@doi [\aap] {10.1051/0004-6361:20065618}, \href
  {https://ui.adsabs.harvard.edu/abs/2006A&A...460..777G} {460, 777}

\bibitem[\protect\citeauthoryear{{Gosset}, {Remy}, {Manfroid}, {Vreux},
  {Balona}, {Sterken}  \& {Franco}}{{Gosset} et~al.}{1991}]{Gosset91}
{Gosset} E.,  {Remy} M.,  {Manfroid} J.,  {Vreux} J.~M.,  {Balona} L.~A.,
  {Sterken} C.,   {Franco} G.~A.~P.,  1991, Information Bulletin on Variable
  Stars, \href {https://ui.adsabs.harvard.edu/abs/1991IBVS.3571....1G} {3571,
  1}

\bibitem[\protect\citeauthoryear{{Hamann}, {Gr{\"a}fener}  \&
  {Liermann}}{{Hamann} et~al.}{2006}]{Hamann06}
{Hamann} W.~R.,  {Gr{\"a}fener} G.,   {Liermann} A.,  2006, \mn@doi [\aap]
  {10.1051/0004-6361:20065052}, \href
  {https://ui.adsabs.harvard.edu/abs/2006A&A...457.1015H} {457, 1015}

\bibitem[\protect\citeauthoryear{{Hamann} et~al.,}{{Hamann}
  et~al.}{2019}]{Hamann19}
{Hamann} W.~R.,  et~al., 2019, \mn@doi [\aap] {10.1051/0004-6361/201834850},
  \href {https://ui.adsabs.harvard.edu/abs/2019A&A...625A..57H} {625, A57}

\bibitem[\protect\citeauthoryear{{Hillier}}{{Hillier}}{2020}]{hillier20}
{Hillier} D.~J.,  2020, \mn@doi [Galaxies] {10.3390/galaxies8030060}, \href
  {https://ui.adsabs.harvard.edu/abs/2020Galax...8...60H} {8, 60}

\bibitem[\protect\citeauthoryear{{Kallrath} \& {Milone}}{{Kallrath} \&
  {Milone}}{2009}]{kall09}
{Kallrath} J.,  {Milone} E.~F.,  2009, {Eclipsing Binary Stars: Modeling and
  Analysis}, \mn@doi{10.1007/978-1-4419-0699-1.
}

\bibitem[\protect\citeauthoryear{{Lamontagne}, {Moffat}, {Drissen}, {Robert}
  \& {Matthews}}{{Lamontagne} et~al.}{1996}]{Lamontagne96}
{Lamontagne} R.,  {Moffat} A. F.~J.,  {Drissen} L.,  {Robert} C.,   {Matthews}
  J.~M.,  1996, \mn@doi [\aj] {10.1086/118175}, \href
  {https://ui.adsabs.harvard.edu/abs/1996AJ....112.2227L} {112, 2227}

\bibitem[\protect\citeauthoryear{{Lepine} \& {Moffat}}{{Lepine} \&
  {Moffat}}{1999}]{LepineMoffat99}
{Lepine} S.,  {Moffat} A. F.~J.,  1999, {Is Clumping Universal in Hot Star
  Winds?}, NOAO Proposal

\bibitem[\protect\citeauthoryear{{L{\'e}pine} \& {Moffat}}{{L{\'e}pine} \&
  {Moffat}}{2008}]{LepineMoffat08}
{L{\'e}pine} S.,  {Moffat} A. F.~J.,  2008, \mn@doi [\aj]
  {10.1088/0004-6256/136/2/548}, \href
  {https://ui.adsabs.harvard.edu/abs/2008AJ....136..548L} {136, 548}

\bibitem[\protect\citeauthoryear{{Martins}, {Schaerer}  \& {Hillier}}{{Martins}
  et~al.}{2005}]{Martins05}
{Martins} F.,  {Schaerer} D.,   {Hillier} D.~J.,  2005, \mn@doi [\aap]
  {10.1051/0004-6361:20042386}, \href
  {https://ui.adsabs.harvard.edu/abs/2005A&A...436.1049M} {436, 1049}

\bibitem[\protect\citeauthoryear{{Moffat} \& {Seggewiss}}{{Moffat} \&
  {Seggewiss}}{1978}]{MoffatSeggwiss78}
{Moffat} A.~F.~J.,  {Seggewiss} W.,  1978, \aap, \href
  {https://ui.adsabs.harvard.edu/abs/1978A&A....70...69M} {70, 69}

\bibitem[\protect\citeauthoryear{{Pablo} et~al.,}{{Pablo}
  et~al.}{2016}]{Pablo16}
{Pablo} H.,  et~al., 2016, \mn@doi [\pasp] {10.1088/1538-3873/128/970/125001},
  \href {https://ui.adsabs.harvard.edu/abs/2016PASP..128l5001P} {128, 125001}

\bibitem[\protect\citeauthoryear{{Parkin} \& {Gosset}}{{Parkin} \&
  {Gosset}}{2011}]{Parkin11}
{Parkin} E.~R.,  {Gosset} E.,  2011, \mn@doi [\aap]
  {10.1051/0004-6361/201016125}, \href
  {https://ui.adsabs.harvard.edu/abs/2011A&A...530A.119P} {530, A119}

\bibitem[\protect\citeauthoryear{{Perrier}, {Breysacher}  \& {Rauw}}{{Perrier}
  et~al.}{2009}]{perrier09}
{Perrier} C.,  {Breysacher} J.,   {Rauw} G.,  2009, \mn@doi [\aap]
  {10.1051/0004-6361/200911707}, \href
  {https://ui.adsabs.harvard.edu/abs/2009A&A...503..963P} {503, 963}

\bibitem[\protect\citeauthoryear{{Pigulski} et~al.,}{{Pigulski}
  et~al.}{2016}]{Pigulski16}
{Pigulski} A.,  et~al., 2016, \mn@doi [\aap] {10.1051/0004-6361/201527872},
  \href {https://ui.adsabs.harvard.edu/abs/2016A&A...588A..55P} {588, A55}

\bibitem[\protect\citeauthoryear{{Popowicz}}{{Popowicz}}{2016}]{Popowicz16}
{Popowicz} A.,  2016, {Image processing in the BRITE nano-satellite mission}.
p. 99041R, \mn@doi{10.1117/12.2229141}

\bibitem[\protect\citeauthoryear{{Popowicz} et~al.,}{{Popowicz}
  et~al.}{2017}]{Popowicz17}
{Popowicz} A.,  et~al., 2017, \mn@doi [\aap] {10.1051/0004-6361/201730806},
  \href {https://ui.adsabs.harvard.edu/abs/2017A&A...605A..26P} {605, A26}

\bibitem[\protect\citeauthoryear{{Ramiaramanantsoa} et~al.,}{{Ramiaramanantsoa}
  et~al.}{2019}]{Ramia19}
{Ramiaramanantsoa} T.,  et~al., 2019, \mn@doi [\mnras] {10.1093/mnras/stz2895},
  \href {https://ui.adsabs.harvard.edu/abs/2019MNRAS.490.5921R} {490, 5921}

\bibitem[\protect\citeauthoryear{{Rauw}, {Vreux}, {Gosset}, {Hutsemekers},
  {Magain}  \& {Rochowicz}}{{Rauw} et~al.}{1996}]{Rauw96}
{Rauw} G.,  {Vreux} J.~M.,  {Gosset} E.,  {Hutsemekers} D.,  {Magain} P.,
  {Rochowicz} K.,  1996, \aap, \href
  {https://ui.adsabs.harvard.edu/abs/1996A&A...306..771R} {306, 771}

\bibitem[\protect\citeauthoryear{{Schweickhardt}, {Schmutz}, {Stahl},
  {Szeifert}  \& {Wolf}}{{Schweickhardt} et~al.}{1999}]{Schwei99}
{Schweickhardt} J.,  {Schmutz} W.,  {Stahl} O.,  {Szeifert} T.,   {Wolf} B.,
  1999, \aap, \href {https://ui.adsabs.harvard.edu/abs/1999A&A...347..127S}
  {347, 127}

\bibitem[\protect\citeauthoryear{Strutz}{Strutz}{2016}]{LM16}
Strutz T.,  2016, Data Fitting and Uncertainty (2nd edition)

\bibitem[\protect\citeauthoryear{{Todt}, {Sander}, {Hainich}, {Hamann}, {Quade}
   \& {Shenar}}{{Todt} et~al.}{2015}]{todt15}
{Todt} H.,  {Sander} A.,  {Hainich} R.,  {Hamann} W.~R.,  {Quade} M.,
  {Shenar} T.,  2015, \mn@doi [\aap] {10.1051/0004-6361/201526253}, \href
  {https://ui.adsabs.harvard.edu/abs/2015A&A...579A..75T} {579, A75}

\bibitem[\protect\citeauthoryear{{Weiss} et~al.,}{{Weiss}
  et~al.}{2014}]{Weiss14}
{Weiss} W.~W.,  et~al., 2014, \mn@doi [\pasp] {10.1086/677236}, \href
  {https://ui.adsabs.harvard.edu/abs/2014PASP..126..573W} {126, 573}

\bibitem[\protect\citeauthoryear{{Wilson}}{{Wilson}}{1979}]{wils79}
{Wilson} R.~E.,  1979, \mn@doi [\apj] {10.1086/157588}, \href
  {https://ui.adsabs.harvard.edu/abs/1979ApJ...234.1054W} {234, 1054}

\bibitem[\protect\citeauthoryear{{Wilson} \& {Devinney}}{{Wilson} \&
  {Devinney}}{1971}]{wils71}
{Wilson} R.~E.,  {Devinney} E.~J.,  1971, \mn@doi [\apj] {10.1086/150986},
  \href {https://ui.adsabs.harvard.edu/abs/1971ApJ...166..605W} {166, 605}

\bibitem[\protect\citeauthoryear{{van Hamme}}{{van Hamme}}{1993}]{hamme93}
{van Hamme} W.,  1993, \mn@doi [\aj] {10.1086/116788}, \href
  {https://ui.adsabs.harvard.edu/abs/1993AJ....106.2096V} {106, 2096}

\bibitem[\protect\citeauthoryear{{von Zeipel}}{{von Zeipel}}{1924}]{zeip24}
{von Zeipel} H.,  1924, \mn@doi [\mnras] {10.1093/mnras/84.9.702}, \href
  {https://ui.adsabs.harvard.edu/abs/1924MNRAS..84..702V} {84, 702}

\makeatother
\end{thebibliography}

%\begin{thebibliography}{99}
%\end{thebibliography}

%%%%%%%%%%%%%%%%%%%%%%%%%%%%%%%%%%%%%%%%%%%%%%%%%%

%%%%%%%%%%%%%%%%% APPENDICES %%%%%%%%%%%%%%%%%%%%%

%\appendix


%If you want to present additional material which would interrupt the flow of the main paper, it can be placed in an Appendix which appears after the list of references.

%%%%%%%%%%%%%%%%%%%%%%%%%%%%%%%%%%%%%%%%%%%%%%%%%%

% Don't change these lines
\bsp	% typesetting comment
\label{lastpage}
\end{document}